\documentclass[12pt,a4paper]{article}    
\usepackage{jheppub}
\makeatletter
\renewcommand\@fpheader{\color{white}'}

\usepackage[T1]{fontenc}
\usepackage[utf8]{inputenc}
\usepackage[british]{babel}
\usepackage{amssymb,amstext,amsthm,
bbm,mathrsfs,amscd,bbold,pifont, tensor}
\usepackage{mathtools}
\usepackage{array}
\usepackage[svgnames,table]{xcolor}
\usepackage[T1]{fontenc}
\usepackage[utf8]{inputenc}
\usepackage{enumitem}
\usepackage{cleveref}
\crefformat{footnote}{#2\footnotemark[#1]#3}

\usepackage{booktabs,caption}
\usepackage{verbatim}
\usepackage{braket}
\usepackage{mdframed}
\usepackage{tikz,pgfplots}
\usetikzlibrary{calc,arrows,cd}

\hypersetup{%
  pdftitle   = {Null fluid/gravity correspondence},
  pdfkeywords = {ultra-relativistic hydrodynamics, null hydrodynamics, fluid/gravity correspondence, black branes, blackfold, pp-waves, Penrose limits},
  pdfauthor  = {Jay Armas, Emil Have, Gianbattista-Piero Nicosia},
}
\theoremstyle{plain}

\theoremstyle{definition}

\newcommand{\so}{\mathfrak{so}}

\ifdefined\C
\renewcommand{\C}{\boldsymbol{C}}
\else
\newcommand{\C}{\boldsymbol{C}}
\fi

\newcommand{\D}{\partial}

\definecolor{gris}{rgb}{0.5,0.5,0.5}
\definecolor{darkgreen}{rgb}{0.0,0.5,0.0}

\allowdisplaybreaks[0]
\numberwithin{equation}{section}
\usetikzlibrary{shapes,decorations}
\pgfdeclarelayer{nodelayer}
\pgfdeclarelayer{edgelayer}
\pgfsetlayers{edgelayer,nodelayer,main}
\tikzstyle{ghost}=[fill=none, draw=none, shape=circle]
\tikzstyle{cross}=[fill=none, draw=black, shape=circle]
\tikzstyle{grey line}=[-, draw={rgb,255: red,128; green,128; blue,128}]
\tikzstyle{blue line}=[-, draw=blue, fill={rgb,255: red,162; green,246; blue,255}]
\tikzstyle{dash blue}=[-, draw=blue, dashed]
\tikzstyle{dash grey}=[-, draw={rgb,255: red,128; green,128; blue,128}, dashed]
\tikzstyle{thick black line}=[-, thick]
\tikzstyle{blue fill}=[-, draw=none, fill={rgb,255: red,126; green,214; blue,255}]
\tikzstyle{black line}=[-]
\tikzstyle{red fill}=[-, fill={rgb,255: red,255; green,162; blue,164}, draw={rgb,255: red,255; green,0; blue,4}]
\tikzstyle{new edge style 0}=[-, draw=none, fill={rgb,255: red,255; green,162; blue,164}]
\pgfplotsset{compat=1.18}

\begin{document}

\title{
\bf{
Null fluid/gravity correspondence
}}

\author[1,2,3,4]{Jay Armas,}
\author[5,6]{Emil Have,}
\author[1]{Gianbattista-Piero Nicosia}

\affiliation[1]{Institute for Theoretical Physics, University of Amsterdam, \\ 1090 GL Amsterdam, The Netherlands}
\affiliation[2]{Dutch Institute for Emergent Phenomena,\\ 
1090 GL Amsterdam, The Netherlands}
\affiliation[3]{Institute for Advanced Study, University of Amsterdam,\\ 
Oude Turfmarkt 147, 1012 GC Amsterdam, The Netherlands}
\affiliation[4]{Niels Bohr International Academy, The Niels Bohr Institute,\\
University of Copenhagen, Blegdamsvej 17, DK-2100 Copenhagen \O, Denmark}
\affiliation[5]{The Mathematical Institute, University of Oxford,\\ Woodstock Road, Oxford OX2 6GG, United Kingdom}
\affiliation[6]{Center of Gravity, Niels Bohr Institute, University of Copenhagen,\\
Blegdamsvej 17, DK-2100 Copenhagen Ø, Denmark}


\emailAdd{j.armas@uva.nl}
\emailAdd{emil.have@maths.ox.ac.uk}
\emailAdd{g.nicosia@uva.nl}

\abstract{We construct a new class of perturbative asymptotically Anti-de Sitter pp-wave spacetimes by performing a long-wavelength expansion of Kaigorodov metrics in arbitrary spacetime dimensions. Holographically, these spacetimes are described by a null fluid hydrodynamic expansion around null states in the conformal field theory, which can be obtained as zero temperature and infinite momentum limits of finite temperature states. Building on this, we explicitly show that special cases of this null fluid/gravity correspondence can be obtained as an ultra-relativistic limit of the usual fluid/gravity correspondence in which the temperature tends to zero while the flow approaches the speed of light. We also extend these results to the context of the blackfold approach in which the corresponding pp-wave spacetimes are asymptotically flat and can be obtained as infinite temperature limits of boosted black branes.}

\maketitle

\section{Introduction}
\label{sec:introduction}
The fluid/gravity correspondence~\cite{Bhattacharyya:2007vjd,Rangamani:2009xk,Hubeny:2011hd} can be understood as a classical, long-wavelength realisation of the AdS/CFT correspondence that exploits a separation of scales to relate solutions of Einstein's equations in a $(D = d+1)$-dimensional asymptotically anti-de Sitter bulk to hydrodynamic flows on the $d$-dimensional conformal boundary.  
In this regime, a set of the bulk equations reduces to the conservation of the boundary energy-momentum tensor, giving rise to the hydrodynamic equations of motion. 
Since its original formulation \cite{Bhattacharyya:2007vjd}, based on slowly varying deformations of a neutral AdS$_5$ black brane, the correspondence has been extended systematically in various directions, e.g., to higher dimensions~\cite{Bhattacharyya:2008mz,Haack:2008cp}, by including conserved charges and gauge anomalies~\cite{Erdmenger:2008rm,Banerjee:2008th}, gravitational anomalies \cite{Megias:2013joa}, higher-form charges \cite{Grozdanov:2018fic, Armas:2019sbe, Davison:2025sze}, to mention only a few.\footnote{See also~\cite{Arenas-Henriquez:2025rpt} for a discussion of fluid/gravity for AdS Robinson--Trautman spacetimes, which are related to AdS-pp-waves.} This has enabled a consistent identification of first- and second-order transport coefficients of the dual relativistic fluid via bulk gravitational calculations~\cite{Baier:2007ix}. As such, the correspondence provides a powerful tool to probe the long-wavelength sector of strongly coupled gauge theories.

The existence of a separation of scales is usually associated with finite temperature physics. Characteristic perturbation wavelengths are chosen to be much larger than the mean free path, typically related to an inverse power of the temperature. In the context of the fluid/gravity correspondence, the temperature of certain states in the conformal field theory (CFT) is identified with the temperature of the black hole horizon while the flow velocity in the CFT is related to the boost parameters of the black hole metric. Is it possible to find such separation of scales in other classes of CFT states and corresponding gravity duals without introducing any extra structure such as additional conserved charges? Einstein gravity with a cosmological constant admits other classes of “planar-like” solutions besides planar AdS black holes. One such class, first found in \cite{Cvetic:1998jf}, consists of families of plane-fronted waves (or pp-waves), in particular, Kaigorodov spacetimes which preserve 1/4 of the supersymmetries \cite{Brecher:2000pa, Gauntlett:2003fk}. From a holographic point of view, these metrics correspond to CFT states whose expectation value of the energy-momentum tensor carries a null momentum density~\cite{Brecher:2000pa}. A subset of these states may be obtained as a double-scaling limit of thermal states, in which the boost is sent to infinity while the temperature goes to zero.\footnote{In earlier literature, Kaigorodov spacetimes, which can be obtained from double-scaling zero temperature limits of planar AdS black holes \cite{Singh:2010zs, Narayan:2012wn}, were interpreted as describing a CFT in an infinite momentum frame \cite{Cvetic:1998jf, Brecher:2000pa} with a finite momentum density stress tensor. These double scaling limits are different from Penrose limits.}  Part of the goal of this paper is to show that such metrics admit a long-wavelength expansion, which from the CFT point of view corresponds to the null fluid expansion recently developed in \cite{Armas:2025uyv}. This fluid expansion is organised by a local scale $\kappa$, which for null fluids plays the same role as the temperature does in ordinary fluids. When the null fluid is obtained from a double-scaling limit, $\kappa$ is the remaining finite scale associated with the pp-wave amplitude, while in the intrinsic description, $\kappa$ fixes the normalisation of the null fluid velocity.

Another goal of the paper is related to a close relative of the fluid/gravity correspondence, namely the blackfold approach~\cite{Emparan:2009cs,Emparan:2009at}, which provides an effective long-wavelength description of asymptotically flat black branes (and more general higher-dimensional black objects). In this framework, one perturbs a black $p$-brane on scales much larger than its thickness, obtaining an effective $(p+1)$-dimensional worldvolume theory whose intrinsic equations take the form of hydrodynamics and whose validity is ensured by solving the bulk Einstein equations order-by-order in gradients~\cite{Camps:2010br}. While the organising principles are similar to fluid/gravity, an important difference is that blackfold dynamics is not formulated in terms of conformal boundary data of an asymptotically AdS spacetime; rather, it is a worldvolume effective theory derived via matched asymptotic expansions, and it naturally incorporates extrinsic (bending/embedding) degrees of freedom~\cite{Camps:2012hw,Armas:2011uf,Armas:2013aka}, which are absent in the standard fluid/gravity setup.\footnote{For a comparison of the two approaches in the case of D3 branes see~\cite{Emparan:2013ila}.} The blackfold method has been applied widely to study the phase structure of higher-dimensional black holes in asymptotically flat and (A)dS settings, as well as in supergravity backgrounds such as AdS$_l\times S^m$ (including charged configurations)~\cite{Emparan:2011hg,Armas:2015nea,Armas:2024dtq,Grignani:2010xm,Grignani:2011mr}, revealing a rich landscape of horizon topologies and associated phases of black holes. In this context, we also consider the analogue flat space metrics of Kaigorodov spacetimes that can be obtained as a double scaling limit of asymptotically flat black branes in which the temperature tends to infinity \cite{Singh:2012un, Armas:2025uyv}. We will show that such spacetimes also admit a hydrodynamic expansion.

\paragraph{Outline.} This paper is organised as follows: in Section \ref{sec:nullhydrodynamics} we provide a review of null hydrodynamics as developed in the recent work~\cite{Armas:2025uyv} by the authors, starting in Section~\ref{sec:nullfluids} with a general discussion of the theory, before focusing on the constant pressure sub-sector, which is the sector relevant in the context of fluid/gravity, and the lightlike (or null) limit in Sections~\ref{sec:const-pressure-subsector} and~\ref{sec:nulllimitofrelfluids}, respectively. 

This is followed by Section~\ref{sec:asyadsnullsol}, where we study hydrodynamic perturbations of AdS-pp-wave spacetimes. In Section~\ref{sec:idealads}, we show how the null limit of a boosted AdS black brane leads to AdS-pp-wave (or Kaigorodov) spacetimes, and how the associated limit of the Brown--York prescription produces the energy-momentum tensor of a pressureless null fluid. Then, in Section~\ref{sec:AdSbottomup}, we consider first-derivative corrections to the ideal-order AdS-pp-wave geometry using the techniques of fluid/gravity. This is followed by a discussion of hydrodynamic frame choices in the null limit in Section~\ref{sec:landauframefromnulllimit}. 

Section~\ref{sec:asyflatnullsol} is the asymptotically flat counterpart of Section~\ref{sec:asyadsnullsol}: we take the null limit of black $p$-branes in Section~\ref{sec:idealflat} to get pp-wave spacetimes and demonstrate that the by Brown--York tensor again produces a pressureless null fluid, while first-derivative corrections to the pp-waves obtained as limits are considered in Section~\ref{sec:firstflat}. Then, in Section~\ref{sec:nullflat} we again discuss hydrodynamic frame choices in the null limit.

We conclude with a brief summary and a discussion of future directions in Section~\ref{sec:discussion}. The manuscript includes three appendices: Appendix~\ref{app:firstordermetrics} reviews the first-order metrics used in fluid/gravity and in the blackfold approach, starting with asymptotically flat blackfold solutions in Section~\ref{app:firstflat}. This is followed by the construction of first-order planar AdS black brane solutions in Schwarzschild-like coordinates using the AdS/Ricci-flat correspondence. In Appendix~\ref{app:EFEs-for-fmunu}, we write down the explicit components of Einstein's equations that we solve to obtain the fluid metrics, with AdS-pp-waves considered in Section~\ref{app:explicitEFE}, and pp-waves considered in Section~\ref{app:dynamicalflat}. Finally, Appendix~\ref{app:frameandcoord} collects a number of technical results regarding frame and coordinate transformations of the fluid metrics.

\paragraph{Notation.} We work with $D$-dimensional Lorentzian manifolds coordinatised by $x^\mu$, where $\mu,\nu,\dots = 0,\dots,D-1$. We write the metric as $g_{\mu\nu}$ and its Levi--Civita connection $\nabla_\mu$.  In the AdS fluid/gravity part, the bulk dimension is $D=d+1$ and the conformal boundary is $d$-dimensional. Boundary coordinates are denoted $\sigma^a$ with Latin indices $a,b,\ldots=0,\ldots,d-1$, and the boundary metric is the flat Minkowski metric $\eta_{ab}$. In the asymptotically flat blackfold part, we consider black $p$-branes in $D=p+n+3$ dimensions, where $p+1$ is the worldvolume dimension and $n+2$ counts the transverse spatial directions. Worldvolume coordinates are again denoted $\sigma^a$, now with $a,b,\ldots=0,\ldots,p$, and the worldvolume metric is $\eta_{ab}$. We reserve the indices $i,j,\ldots$ from the middle of the Latin alphabet for directions transverse to a chosen null direction on the worldvolume when working in adapted coordinates. Finally, we use the shorthand symbol ``$\rightsquigarrow$'' to denote the null (ultra-relativistic) limit of quantities. For example, if a quantity $f(\gamma,T)$ has a well-defined limit in a specified scaling regime $(\gamma,T)\to(\infty,T_L)$, we write
\begin{equation}
	f(\gamma,T)\rightsquigarrow f_{\text{null}}\,,
\end{equation}
with $f_{\text{null}}$ the resulting finite null datum.

\section{Effective theory for null matter}
\label{sec:nullhydrodynamics}

In this section, we review the effective theory of fluids moving at the speed of light following the construction of~\cite{Armas:2025uyv}. This theory can be formulated from first principles by introducing a null fluid velocity vector and implementing a gradient expansion. A special class of these null fluids, which we also discuss below, can be obtained by a double scaling limit of a timelike fluid, in which the fluid velocity is infinitely boosted while the temperature approaches a limiting value.

\subsection{Null hydrodynamics with explicitly broken boosts}
\label{sec:nullfluids}

A null fluid is characterised by a fluid velocity that is null. Geometrically, the kinematic data describing such a null fluid is a null congruence: let $v^\mu$ be a nowhere-vanishing null vector in a $(d+1)$-dimensional Lorentzian spacetime with metric $g_{\mu\nu}$, i.e., $g_{\mu\nu}v^\mu v^\nu = 0$, where $\mu,\nu = 0,\dots,d$. To express the metric in a way that is adapted to $v^\mu$, it is useful to define an auxiliary null vector $\tau^\mu$ satisfying $g_{\mu\nu}\tau^\mu\tau^\nu=0$ and $\tau^\mu v_\mu = -1$. In terms of these objects, components of the metric take the form
\begin{equation}
\label{eq:metric-decomp}
    g_{\mu\nu} = -2v_{(\mu} \tau_{\nu)} + h_{\mu\nu}\,,\qquad h_{\mu\nu} = e^A_\mu e^B_\nu\delta_{AB}\,,
\end{equation}
where we wrote the corank-2 spatial projector $h_{\mu\nu}$ in terms of vielbeine $e^A_\mu$ with $A=1,\dots,d-1$ satisfying $e_\mu^A v^\mu = 0 = e^A_\mu \tau^\mu$, along with $g^{\mu\nu} e_\mu^A e^B_\nu = \delta^{AB}$. The infinitesimal transformations that preserve the form of the metric in~\eqref{eq:metric-decomp} are given by the stabiliser of the null direction provided by $v^\mu$ in the local Lorentz algebra, and they act as
\begin{equation}
\label{eq:infinitesimal-transf}
    \delta v^\mu = \alpha v^\mu\,,\ \ \ \delta\tau_\mu = -\alpha\tau_\mu + \lambda_\mu\,,\ \ \ \delta e^A_\mu =  \lambda^A v_\mu + O^A{_B}e^B_\mu\,,\ \ \ \delta h_{\mu\nu} = 2\lambda_{(\mu}v_{\nu)}\,,
\end{equation}
where $O^A{_B} \in \so(d-1)$ is an infinitesimal rotation, while $\lambda_\mu = \lambda_A e^A_\mu$ where $\lambda_A$ parametrises Lorentz boosts in the $(\tau_\mu,e^A_\mu)$ plane, known as null rotations, and $\alpha$ parametrises the remaining null boost in the $(v^\mu,\tau_\mu)$ plane. In other words, specifying a null vector $v^\mu$ reduces the local Lorentz group to the subgroup that preserves the null direction it spans. In this sense, $v^\mu$ plays the same role as the timelike fluid velocity does for an ordinary fluid. 
As observed in~\cite{Armas:2025uyv}, because $v^\mu$ is defined only up to local rescalings, if one wishes to formulate an effective theory in terms of these geometric objects, no invariant structures exist unless null boosts parametrised by $\alpha$ are broken. This can be achieved in two ways: either they are broken spontaneously, in which case there is a ``Goldstone-like'' field transforming under $\alpha$, or they are explicitly broken, which amounts to choosing a preferred normalisation for $v^\mu$ and introduces a local scale $\kappa(x)$. In this paper, we will focus on null fluids with explicitly broken null boosts, for which the ideal energy-momentum tensor takes form
\begin{equation}
\label{eq:idealnullT}
    \mathcal{T}^{\mu\nu}_{(0)}=\mathcal{E}(\kappa)v^\mu v^\nu+\mathcal{P}(\kappa)g^{\mu\nu}\,,
\end{equation} 
where $\mathcal{E}$ is the null energy density and $\mathcal{P}$ the null pressure, both parametrised in terms of the local scale $\kappa(x)$, and related by an equation of state 
\begin{equation}
\label{eq:EoS}
\mathcal{P} = \mathcal{P}(\mathcal{E})\,.    
\end{equation}
The null fluid equation of motion, which follows from diffeomorphism invariance, tells us that the energy-momentum tensor $\mathcal{T}^{\mu\nu}$ to all orders in the derivative expansion is covariantly conserved
\begin{equation}
\label{eq:consstress}
    \nabla_\mu\mathcal{T}^{\mu\nu}=0\,,
\end{equation} 
where $\nabla_\mu$ is the Levi-Civita connection of $g_{\mu\nu}$. Projecting~\eqref{eq:consstress} along $v^\mu$, $\tau^\mu$, and $h^{\mu\nu}$ using the ideal order energy-momentum tensor~\eqref{eq:idealnullT}, we obtain the equations of motion
\begin{subequations}
    \label{eq:eomESB}
\begin{align}
 v^\mu\D_\mu \mathcal{P} &=0+ \mathcal{O}(\partial^2)\,, \label{eq:first-ideal-eom}\\
 -\nabla_\mu(\mathcal{E}v^\mu)+\mathcal{E}\tau_\mu \dot{v}^\mu+\tau^\mu \D_\mu \mathcal{P}&=0+\mathcal{O}(\partial^2) \,,\\
\mathcal{E} h_{\alpha\mu}\dot{v}^\mu+h_{\alpha}^\mu\D_\mu \mathcal{P}&=0+\mathcal{O}(\partial^2)\,,
\end{align}
\end{subequations}
where we defined the acceleration\footnote{This acceleration was called $a^\nu$ in~\cite{Armas:2025uyv}.}
\begin{equation}
\label{eq:acc-def}
\dot{v}^\nu := v^\mu\nabla_\mu v^\nu\,.    
\end{equation}
The conservation of the energy-momentum tensor~\eqref{eq:consstress} supplies $(d+1)$ equations for the $(d+1)$ degrees of freedom made up of $v^\mu$ and $\kappa$. 

In close analogy with ordinary hydrodynamics, we include first-derivative corrections by expanding in a small dimensionless parameter controlling gradients along the null congruence. In this regard, $\kappa(x)$ provides a local scale that organises the derivative expansion.\footnote{\label{fn:kappa}In this sense, the scale $\kappa(x)$ plays the role of the scale that replaces the temperature in the derivative expansion in ordinary timelike hydrodynamics. This statement should not be understood thermodynamically, since a null state has no local rest frame. Instead, as discussed above, $\kappa(x)$ sets the local null scale by picking out a preferred normalisation of the null fluid velocity. } Concretely, if $\mathcal{R}$ denotes the characteristic length scale over which $\kappa$ and $v^\mu$ vary, we assume that $\D/\kappa \sim \ell_{\text{mfp}}/\mathcal{R} \ll 1$, and write the most general first-order stress tensor as
\begin{equation}
\label{eq:firstnullTgen}    \mathcal{T}^{\mu\nu}_{(1)}=\left(\rho_1\vartheta\!+\!\rho_6\frac{v^\alpha\partial_\alpha\kappa}{\kappa}\right) g^{\mu\nu}+2v^{(\mu}\left(\rho_2\vartheta v^{\nu)}+\!\rho_3\dot{v}^{\nu)}+\!\rho_4v^{\nu)}\frac{v^\alpha\partial_\alpha\kappa}{\kappa}+\!\rho_5\frac{\partial^{\nu)}\kappa}{\kappa}\right)-2\eta\varsigma^{\mu\nu}\ ,
\end{equation} 
where all transport coefficients $\rho_1,\dots,\rho_6,\eta$ are functions of $\kappa$. In the above, $\eta$ is the null shear viscosity with $\varsigma^{\mu\nu}:= \nabla^{(\mu}v^{\nu)}$ the null shear tensor, and $\vartheta = \nabla_\mu v^\mu$ the null fluid expansion.
In the general case, where $\mathcal P$ is not constant, we can use the equations of motion~\eqref{eq:eomESB} to eliminate terms with derivatives of $\kappa$, while frame transformations, which are redefinitions of the hydrodynamic data of the form $\kappa\to\kappa+\bar\delta\kappa$ and $v^\mu\to v^\mu+\bar\delta v^\mu$ with $v_\mu\bar\delta v^\mu=0$ and $\bar\delta\kappa\sim \mathcal{O}(\partial)$, $\bar\delta v^\mu\sim \mathcal{O}(\D)$, allow us to bring the first-order energy-momentum tensor to the form
\begin{equation}
\label{eq:firstnullT}
    \mathcal{T}_{(1)}^{\mu\nu}=-2\eta\varsigma^{\mu\nu}\,,
\end{equation} 
which can be understood as the analogue of the Landau frame for null fluids. The equations of motion at first order are then obtained by inserting $\mathcal{T}^{\mu\nu}=\mathcal{T}^{\mu\nu}_{(0)}+\mathcal{T}^{\mu\nu}_{(1)}$ in~\eqref{eq:consstress}, though we refrain from writing them out explicitly.

\subsection{Constant pressure}
\label{sec:const-pressure-subsector}

An important special case of the null fluid discussed in the previous section is when the null pressure $\mathcal{P}$ is constant, while the null energy density is not. It was shown in~\cite{Armas:2025uyv} that this special class of null fluids appears when taking the ultra-relativistic limit of relativistic fluids, which we discuss in Section~\ref{sec:nulllimitofrelfluids}. The general analysis above changes slightly when the pressure is constant, as we now discuss.

As noted in~\cite{Armas:2025uyv}, when $\mathcal{P}$ is constant, the ideal-order energy-momentum tensor~\eqref{eq:idealnullT} has an ``emergent'' gauge redundancy, since $\mathcal{E}\to\phi(\kappa)^{-2}\mathcal{E}$ and $v^\mu\to\phi(\kappa)v^\mu$ leaves the $\mathcal{T}^{\mu\nu}_{(0)}$ invariant. This redundancy is separate from the null boosts with parameter $\alpha$ appearing in~\eqref{eq:infinitesimal-transf}, though it also manifests itself as a rescaling. This redundancy reduces the number of degrees of freedom at ideal order by one, which is matched by a corresponding reduction in the number of equations of motion: when $\mathcal{P}$ is constant, Eq.~\eqref{eq:first-ideal-eom} is trivially satisfied. When including first-order corrections, this symmetry is lost, and the number of degrees of freedom and the number of equations of motion are again (separately) equal to $d+1$. Later in the manuscript, we will see that the same gauge redundancy is present in the gravity duals to ideal-order null fluids. When the null pressure is constant, the fluid equations at ideal order~\eqref{eq:eomESB} reduce to the pair
\begin{equation}
\label{eq:eomESBP}
    \nabla_\mu(\mathcal{E}v^\mu)=\mathcal{E}\tau_\mu \dot v^\mu+\mathcal{O}(\partial^2)\,, \qquad \mathcal{E}h_{\alpha\mu}\dot v^\mu=0+\mathcal{O}(\partial^2)\,.
\end{equation} 
When first-order corrections are included, the most general energy-momentum tensor is still given by~\eqref{eq:firstnullTgen}, but since the null pressure is constant it can no longer be reduced to the form~\eqref{eq:firstnullT}. More precisely, using frame transformations we can bring the first-order energy-momentum tensor to the form
\begin{equation}
\label{eq:firstnullgenP}
    \mathcal{T}_{(1)}^{\mu\nu}=\left(\rho_1\vartheta+\rho_6\frac{v^\alpha\partial_\alpha\kappa}{\kappa}\right) g^{\mu\nu}+2\rho_5\frac{v^{(\mu}\partial^{\nu)}\kappa}{\kappa}-2\eta\varsigma^{\mu\nu}\,.
\end{equation} 
In contrast to the generic case, where $\mathcal{P}$ depends on $\kappa$, the term proportional to $(d\mathcal{P}/d\kappa)\delta\kappa$ needed to remove the first-order terms proportional to $g^{\mu\nu}$ is absent when $\mathcal{P}$ is constant. Similarly, the term proportional to $\rho_5$ cannot be removed because a frame transformation would require that $v^\mu\partial_\mu\kappa=0$,  which is not necessarily true since~\eqref{eq:first-ideal-eom} is trivially satisfied in the constant-pressure case.

\subsection{The lightlike limit of relativistic fluids}
\label{sec:nulllimitofrelfluids}

Certain special cases of null fluids with explicitly broken null boosts and with constant null pressure discussed above may be obtained by taking an ultra-relativistic limit of a timelike fluid, corresponding to an infinite local boost as discussed in~\cite{Armas:2025uyv}. In this section, we review this procedure: consider a unit normalised timelike fluid vector $u^{\mu}$, $g_{\mu\nu}u^\mu u^\nu = -1$, and write
\begin{equation}
    u^\mu = \gamma(x)\, U^\mu\,, \qquad \gamma(x)=\frac{1}{\sqrt{-\,g_{\mu\nu}U^\mu U^\nu}}\,,
\end{equation}
where $U^\mu$ is an unnormalised timelike vector field. The limit $\gamma(x)\to\infty$ then sends $U^\mu$ to a null vector and implements the local infinite boost; in taking this limit one keeps fixed  the combination $(\varepsilon+P)\gamma^2$ appearing below. The energy-momentum tensor at ideal order of a neutral relativistic fluid is given by~\cite{Kovtun:2012rj}
\begin{equation}
    T^{\mu\nu}_{(0)}=(\varepsilon+P)u^{\mu}u^{\nu}+Pg^{\mu\nu}
    =(\varepsilon+P)\gamma^{2}U^{\mu}U^{\nu}+Pg^{\mu\nu}\,,
\end{equation}
where $\varepsilon=\varepsilon(T)$ and $P=P(T)$ are the energy density and pressure, respectively, and $T$ and $s(T)$ are the temperature and entropy density. The pressure and the energy density satisfy the Euler relation $\varepsilon+P=Ts$. 

We assume that the equation of state admits a scaling regime in which the enthalpy density
$w(T):=\varepsilon(T)+P(T)$ tends to zero as $T\to T_L$ (usually either $T_L=0$ or $T_L=\infty$), in such a way that the combination $w(T)\gamma^2$ remains finite.\footnote{This is superficially reminiscent of Carrollian fluids~\cite{Ciambelli:2018xat,deBoer:2021jej,deBoer:2023fnj,Armas:2023dcz}, or, more generally, framids~\cite{Nicolis:2015sra}. We stress, however, that the ultra-relativistic limit, where the fluid velocity becomes lightlike, is very different from the Carrollian limit, where the speed of light goes to zero.}
At leading order, we define the null limit as the simultaneous limit $(T,\gamma)\to (T_{L},\infty)$ such that
\begin{equation}
\label{eq:defnlim0order}
    (\varepsilon+P)\gamma^2\rightsquigarrow\mathcal{E}(\kappa)\,, \qquad U^\mu\rightsquigarrow v^\mu\,, \qquad P\rightsquigarrow\mathcal{P}\,,
\end{equation}
where ``$\rightsquigarrow$'' denotes the null limit, so that $T^{\mu\nu}_{(0)}\rightsquigarrow\mathcal{T}^{\mu\nu}_{(0)}$ with $\mathcal{P}$ constant. We emphasise that $T_L$ is simply a convenient parametrisation of the scaling regime: any limiting temperature for which~\eqref{eq:defnlim0order} holds leads to the same null constitutive data. For example, for asymptotically flat black branes the null limit is realised in the high-temperature regime $(T_L \to \infty)$, while for asymptotically AdS black branes it is realised in the low-temperature regime $(T_L\to 0)$ as we will show in the next sections. Finally, this ideal-order limit fixes the relative scaling of $T$ and $\gamma$, namely 
\begin{equation}
\label{eq:T-gamma-scaling}
T(x)\sim \kappa(x)\gamma^\beta\,,
\end{equation}
with $\beta$ a case-dependent constant, and consequently constrains the scaling of transport coefficients order by order in the derivative expansion.

Proceeding to first order in gradients, the viscous correction to the stress tensor of a neutral relativistic fluid in the Landau frame\footnote{Recall that ``Landau frame'' means that the timelike fluid velocity $u^\mu$ is an eigenvector of the full stress tensor, implying that there is no energy flux in the local rest frame; or, equivalently, 
\begin{equation*}
u_\mu T^{\mu\nu}=-\varepsilon u^\nu
\qquad\iff\qquad
u_\mu T^{\mu\nu}_{(1)}=0\,,
\end{equation*}
where $T^{\mu\nu}=T^{\mu\nu}_{(0)}+T^{\mu\nu}_{(1)}+\cdots$.} can be written as~\cite{Kovtun:2012rj}
\begin{equation}
	\label{eq:T1nonull}
	T^{\mu\nu}_{(1)}=\left(\zeta-\frac{\hat\eta}{d}\right)\left(g^{\mu\nu}+\gamma^{2}U^{\mu}U^{\nu}\right)\theta
	+\hat\eta\left(\nabla^{(\mu}\left(\gamma U^{\nu)}\right)+\gamma^{2}U^{\alpha}U^{(\mu}\nabla_{\alpha}\left(\gamma U^{\nu)}\right)\right)\,,
\end{equation}
where $\zeta(T)$ and $\hat\eta(T)$ are the bulk and shear viscosities, and
$\theta=\nabla_{\mu}u^{\mu}=\nabla_\mu(\gamma U^\mu)$ the timelike fluid expansion. In contrast to ideal order, defining the ultra-relativistic limit at first order requires specifying how the derivative expansion is taken in tandem with $\gamma\to\infty$.

Let $\ell_{\text{mfp}}(T)$ denote the microscopic length scale controlling the validity of hydrodynamics, and let $\mathcal R$ denote the macroscopic scale over which the hydrodynamic fields vary, so that schematically $\nabla\sim \mathcal R^{-1}$. Hydrodynamics requires $\ell_{\text{mfp}}/\mathcal R\ll 1$. The ultra-relativistic scaling~\eqref{eq:T-gamma-scaling} implies that the quantity $\ell_{\text{mfp}}(T)$ generally becomes singular (for instance, $\ell_{\text{mfp}}\propto 1/T$ diverges as $T\to0$), so maintaining $\ell_{\text{mfp}}/\mathcal R \ll 1$ forces a simultaneous scaling of the macroscopic length scale $\mathcal{R}$ with $\gamma$. We parametrise this double scaling by introducing an exponent $\alpha$ via\footnote{This $\alpha$ has not to be confused with the $\alpha$ parametrising null boosts.}
\begin{equation}
	\label{eq:alpha-def}
	\nabla  \sim \mathcal{O}(\gamma^{\alpha})\iff
	\mathcal R \sim \mathcal{O}(\gamma^{-\alpha})\,,
\end{equation}
so that keeping $\ell_{\mathrm{mfp}}\nabla$ small corresponds to
\begin{equation}
	\label{eq:mfp-scaling}
	\ell_{\text{mfp}}\left(T(\gamma)\right) \sim \mathcal{O}(\gamma^{-\alpha})\,.
\end{equation}
In this sense the exponent $\alpha$ is not an independent microscopic input, since once the ideal-order scaling~\eqref{eq:T-gamma-scaling} is fixed, the temperature dependence of $\ell_{\mathrm{mfp}}(T)$ determines the compatible $\alpha$ required for a well-defined derivative expansion.\footnote{Throughout we assume $\partial_\mu\log\gamma$ is subleading, i.e., $\gamma$ varies on scales longer than the hydrodynamic fields (see~\cite{Armas:2025uyv} for more details).}

With the scaling~\eqref{eq:alpha-def}, the first-derivative structures appearing in~\eqref{eq:T1nonull} scale as
\begin{equation}
	\label{eq:scaling-structures-alpha}
	\theta=\nabla_\mu(\gamma U^\mu)\sim \mathcal{O}(\gamma^{\alpha+1})\,,\qquad
	\nabla^{(\mu}(\gamma U^{\nu)})\sim \mathcal{O}(\gamma^{\alpha+1})\,,\qquad
	U^\alpha\nabla_\alpha(\gamma U^\mu)\sim \mathcal{O}(\gamma^{\alpha+1})\,.
\end{equation}
In addition, different tensor structures may involve additional explicit factors of the velocity, leading to additional explicit powers of $\gamma$.
It is convenient to keep track of these by an integer $h$, defined as the number of explicit factors of $u^\mu$ multiplying a given first-derivative structure (for example, a term of the schematic form $X(T) u^\mu u^\nu\,\theta$ has $h=2$).
A generic first-order term then scales as
\begin{equation}
	\label{eq:defnlimit1}
	X(T)\gamma^{\alpha+1+h} \rightsquigarrow Y(\kappa)\,,\qquad \text{where}\qquad  \gamma^{-1}\nabla_\mu u_\nu \rightsquigarrow \nabla_\mu v_\nu\,,
\end{equation}
where $X(T)$ is a timelike transport coefficient (or fixed linear combination thereof) with $T$ scaling as in~\eqref{eq:T-gamma-scaling}, and $Y(\kappa)$ denotes the corresponding finite null transport data.
Equation~\eqref{eq:defnlimit1} is the precise statement of what it means for the null limit to exist at first-order in the derivative expansion. We can now apply this to~\eqref{eq:T1nonull}. We see that the terms involving $\gamma^2 U^\mu U^\nu$ and $\gamma^2 U^\alpha U^{(\mu}\nabla_\alpha(\gamma U^{\nu)})$ correspond to $h=2$ and therefore scale as $\mathcal{O}(\gamma^{\alpha+3})$. Taking the ultra-relativistic limit directly in Landau frame, finiteness requires
\begin{equation}
	\label{eq:limitT1landau}
	\left(\zeta-\frac{\hat\eta}{d}\right)\gamma^{\alpha+3}\rightsquigarrow 2\rho_2(\kappa)\,,\qquad
	\hat\eta \gamma^{\alpha+3}\rightsquigarrow 2\rho_3(\kappa)\,,
\end{equation}
so that, using $U^\mu\rightsquigarrow v^\mu$, one obtains
\begin{equation}
	\label{eq:limitT1landau-result}
	T^{\mu\nu}_{(1)} \rightsquigarrow 2\rho_2 v^\mu v^\nu \vartheta + 2\rho_3 v^{(\mu}\dot v^{\nu)}\,,
\end{equation}
where $\dot v^\mu$ and $\vartheta$ are defined in~\eqref{eq:acc-def} and~\eqref{eq:firstnullTgen}, respectively. Comparing with~\eqref{eq:firstnullTgen}, these are indeed admissible first-order structures in the most general null constitutive relation, but, as explained in Section~\ref{sec:nullfluids}, these terms can be removed by a suitable frame transformation. As discussed in~\cite{Armas:2025uyv}, we may perform an ordinary timelike frame transformation before taking the limit, i.e., $u^\mu\to u^\mu+\bar\delta u^\mu$ and $T\to T+\bar\delta T$ with $\bar\delta u^\mu,\bar\delta T \sim \mathcal{O}(\D)$. In this non-thermodynamic frame one may write~\eqref{eq:T1nonull} equivalently as
\begin{equation}
	\label{eq:nullLandauT1}
	T^{\mu\nu}_{(1)}=\hat\zeta \theta  g^{\mu\nu}-\hat\eta \nabla^{(\mu}u^{\nu)}\,,
	\qquad
	\hat\zeta=\left(\zeta-\frac{\hat\eta}{d}\right)\left(1-\frac{s}{\partial(Ts)/\partial T}\right)\,,
\end{equation}
where the explicit form of $\hat\zeta$ follows from the chosen frame redefinition.
In this representation the relevant terms have $h=0$, so the existence of a finite null limit requires
\begin{equation}
	\label{eq:limitT1nulllandu}
	\hat\zeta\gamma^{\alpha+1}\rightsquigarrow \rho_1(\kappa)\,,\qquad
	\hat\eta\gamma^{\alpha+1}\rightsquigarrow 2\eta(\kappa)\,,
\end{equation}
leading to
\begin{equation}
	\label{eq:limitT1nulllandu-result}
	T^{\mu\nu}_{(1)} \rightsquigarrow \rho_1 \vartheta g^{\mu\nu}-2\eta\,\varsigma^{\mu\nu}\,.
\end{equation}
In this way, the null limit in the non-thermodynamic frame produces a subset of the general constant-pressure energy-momentum tensor~\eqref{eq:firstnullgenP}.

If present, conformal symmetry imposes further constraints on the timelike constitutive relations~\cite{Kovtun:2019hdm}, and one must ensure that any frame transformation respects these constraints. A convenient representation of the first-order stress tensor for a conformal fluid is
\begin{equation}
	\label{eq:newnonthermoT}
	T^{\mu\nu}_{(1)}=\left(-\frac{\hat\eta}{d+1}\frac{u^\alpha\partial_\alpha T}{T}+\frac{\hat\eta}{d+1}\theta \right)g^{\mu\nu}+\hat\eta\,\frac{\partial^{(\mu}T}{T}u^{\nu)}-\hat\eta\nabla^{(\mu}u^{\nu)}\,.
\end{equation}
Using the ideal-order scaling~\eqref{eq:T-gamma-scaling}, together with $\partial_\mu\log\gamma\to0$, we have $\partial_\mu T/T \rightsquigarrow \partial_\mu\kappa/\kappa$.
Taking the null limit with the same $(T,\gamma)\to(T_L,\infty)$ scaling and demanding that 
\begin{equation}
\label{eq:Xi}
\hat\eta(T) \gamma^{\alpha+1} \rightsquigarrow \{\rho_{1,5,6}(\kappa),\eta(\kappa)\}\,,    
\end{equation}
is finite in the limit yields
\begin{equation}
	\label{eq:limit-conformal}
	T^{\mu\nu}_{(1)}\rightsquigarrow
	\left(\rho_1\vartheta+\rho_6 \frac{v^\alpha\partial_\alpha\kappa}{\kappa}\right) g^{\mu\nu}+2\rho_5\,\frac{v^{(\mu}\partial^{\nu)}\kappa}{\kappa}-2\eta\varsigma^{\mu\nu}\,,
\end{equation}
which matches the most general constant-pressure null constitutive relation~\eqref{eq:firstnullgenP} with the identifications
\begin{equation}
	\label{eq:rho-identifications}
	\rho_1=\frac{\hat\eta\gamma^{\alpha+1}}{d+1}\,,\qquad \rho_5=\frac{\hat\eta \gamma^{\alpha+1}}{2}\,,\qquad \rho_6=-\frac{\hat\eta \gamma^{\alpha+1}}{d+1}\,,\qquad \eta=\frac{\hat\eta \gamma^{\alpha+1}}{2}\,.
\end{equation}
The resulting null energy-momentum tensor is automatically traceless and the same relations will be reproduced by the gravitational calculation.

We have thus shown how null limits can be defined such that we recover (special cases) of the null energy-momentum tensor. In particular, the detailed form of this energy-momentum tensor obtained in the limit is highly dependent on the frame of the timelike fluid. One can explicitly check that taking the most general frame for a timelike fluid~\cite{Kovtun:2019hdm} as the starting point either gives \eqref{eq:limitT1landau-result} in the null limit but with an additional independent $\rho_4$ coefficient or precisely lands on the form of~\eqref{eq:limit-conformal} with 4 independent coefficients. It is not possible, by means of a null limit of a timelike fluid, to obtain the most general null fluid in~\eqref{eq:firstnullTgen}. As we will see below, this also happens on the gravitational side: the null limit of hydrodynamically perturbed black branes reproduces only the branch of solutions that arise as limits of timelike fluids. To obtain the most general asymptotically Kaigorodov solutions and asymptotically flat null brane solutions, we must consider long-wavelength perturbations around a null background and solve the corresponding Einstein equations.

\section{Asymptotically AdS null solutions}
\label{sec:asyadsnullsol}

In this section we construct asymptotically AdS bulk solutions of Einstein's equations in a null hydrodynamic derivative expansion and extract their boundary stress tensors, obtaining a conformal viscous null fluid. At ideal order the bulk geometry arises as the ultra-relativistic limit of a boosted AdS black brane. We then implement the fluid/gravity procedure at first order, solving the Einstein equations for slowly varying $\kappa(x)$ and null fluid velocity, and show that the resulting metrics agree with those obtained by taking the null limit of the corresponding timelike fluid/gravity solution. We have checked the construction explicitly for $d=2,3,4,5$ and conjecture it holds for general $d$.

\subsection{Ideal order from null limit}
\label{sec:idealads}

Einstein's equations in $D=d+1$ bulk dimensions with negative cosmological constant $\Lambda$ are given by
\begin{equation}
	\label{eq:Edefn}
	E_{\mu\nu}:=R_{\mu\nu}-\frac12\,(R-2\Lambda)\,g_{\mu\nu}=0\,.
\end{equation}
We split bulk coordinates as $x^\mu=(r,\sigma^a)$, where $r$ is the radial coordinate and $\sigma^a$ ($a=0,\dots,d-1$) are boundary coordinates. In units where the AdS radius is set to one, $\Lambda=-\frac{1}{2} d(d-1)$, the boosted AdS black brane metric may be written as
\begin{equation}
	\label{eq:dsSAds}
	ds^2=\frac{dr^2}{r^2 f(br)}+r^2\Big(\eta_{ab}-(f(br)-1)u_a u_b\Big)d\sigma^a d\sigma^b\,,
	\qquad
	f(br)=1-\frac{1}{(rb)^{d}}\,,
\end{equation}
where $\eta_{ab}$ is the flat boundary metric, $u^a$ is a unit timelike velocity field ($\eta_{ab}u^a u^b=-1$), and $b$ is a (constant) length scale so that $rb$ is dimensionless.

The holographic energy-momentum tensor associated with an asymptotically AdS solution is obtained from the renormalised Brown--York tensor~\cite{Balasubramanian:1999re}. For a flat boundary metric, the required counterterm reduces to the cosmological term, and the boundary energy-momentum tensor can be written as
\begin{equation}
	\label{eq:AdSTabdefn}
	T_{ab}=\frac{1}{8\pi G_{D}}\lim_{r\to\infty}\Big[r^{d-2}\big(K_{ab}-K\gamma_{ab}-(d-1)\gamma_{ab}\big)\Big]\,,
\end{equation}
where $\gamma_{ab}$ is the induced metric on a constant-$r$ hypersurface and $K_{ab}$ its extrinsic curvature with trace $K=\gamma^{ab}K_{ab}$, while $G_D$ is  the $D$-dimensional gravitational constant. Applied to~\eqref{eq:dsSAds}, this yields the ideal conformal fluid stress tensor
\begin{equation}
	\label{eq:TabAdSideal}
	T^{(0)}_{ab}=P\big(\eta_{ab}+du_a u_b\big)\,,
	\qquad
	P=\frac{1}{16\pi G_{D}b^{d}}\,.
\end{equation}
The corresponding temperature and entropy density read
\begin{equation}
	\label{eq:TsAdS}
	T=\frac{d}{4\pi b}\,,
	\qquad
	s=\frac{1}{4G_{D}b^{d-1}}\,,
\end{equation}
so that thermodynamic quantities may be parametrised equivalently by $T$ or by $b$.

We now obtain the ideal-order AdS null geometry by taking the ultra-relativistic limit discussed in Section~\ref{sec:nulllimitofrelfluids}. Writing the unit timelike velocity as $u^a=\gamma U^a$, with $U^a$ an unnormalised timelike vector field as above, this null limit is defined by
\begin{equation}
	\label{eq:null-limit-b}
	\frac{\gamma^{2}}{b^{d}} \rightsquigarrow \kappa^{d}\,,
	\qquad
	U_a \rightsquigarrow v_a\,,
\end{equation}
so that $\kappa$ has dimensions $[\kappa]=[b]^{-1}$ and the temperature $T\propto b^{-1}$ tends to zero in this scaling regime. Taking the limit of~\eqref{eq:dsSAds} then gives\footnote{This AdS-pp-wave may be viewed as the planar, asymptotically locally AdS analogue of the AdS shock waves obtained by boosting AdS black holes at fixed energy to the speed of light, which were considered in~\cite{Horowitz:1999gf}.}
\begin{equation}
	\label{eq:dsAdSnullideal}
	ds^2=\frac{dr^{2}}{r^{2}}+r^{2}\left(\eta_{ab}+\frac{\kappa^{d}}{r^d}v_{a}v_{b}\right)d\sigma^a d\sigma^b\,,
\end{equation}
which is an AdS-pp-wave (a Kaigodorov spacetime \cite{Kaigorodov, Cvetic:1998jf}) which is a special class of a Siklos metric \cite{Siklos,Podolsky:1997ik}. It is asymptotically (locally) AdS in the usual Fefferman--Graham sense~\cite{deHaro:2000vlm,Skenderis:2002wp}, but it no longer carries a finite-temperature horizon. Moreover, as is typical for AdS plane-wave geometries, scalar curvature invariants coincide with those of pure AdS; for instance the Kretschmann scalar is
\begin{equation}
	R^{\alpha\beta\gamma\delta}R_{\alpha\beta\gamma\delta}=2d(d+1)\,,
\end{equation}
in our units where the AdS radius has been set to one.

We furthermore note that $\kappa$ enters the AdS-pp-wave geometry only through the rank-1 deformation of the flat background 
\begin{equation}
	g_{ab}(r,\sigma)=r^{2}\eta_{ab}+r^{2-d}\,\kappa(\sigma)^d\,v_a(\sigma)v_b(\sigma)\,.
\end{equation}
At ideal order the combination $\kappa^{d}v_a v_b$ is invariant under the local rescaling
\begin{equation}
	\label{eq:emergent-null-gauge}
	v_a(\sigma) \to e^{\lambda(\sigma)}v_a(\sigma)\,,
	\qquad
	\kappa(\sigma) \to e^{-2\lambda(\sigma)/d}\kappa(\sigma)\,,
\end{equation}
which leaves the metric unchanged. In other words, the bulk ansatz does not depend on $\kappa$ and $v_a$ separately, but only on the gauge-invariant composite object $\kappa^{d}v_a v_b$. Equivalently, one may introduce the invariant null one-form
\begin{equation}
	\ell_a:=\kappa^{d/2}v_a\,,\qquad \ell^2=0\,,
\end{equation}
in terms of which the metric takes the manifestly gauge-invariant form
\begin{equation}
	ds^2=\frac{dr^{2}}{r^{2}}+r^{2}\left(\eta_{ab}+\frac{1}{r^d}\,\ell_a\ell_b\right)d\sigma^a d\sigma^b\,.
\end{equation}
This emergent rescaling symmetry is the bulk counterpart of the ideal-order scaling redundancy of the null fluid discussed in Section~\ref{sec:nullfluids}: it removes one degree of freedom at ideal order by identifying different pairs $(\kappa,v_a)$ that correspond to the same spacetime geometry. By construction, the metric~\eqref{eq:dsAdSnullideal} solves Einstein's equations with $\Lambda<0$.

We now turn to the boundary energy-momentum tensor of the ideal-order geometry. Since the boosted AdS black brane~\eqref{eq:dsSAds} has pressure $P\propto b^{-d}$ and temperature $T\propto b^{-1}$, taking the null limit requires $b \to \infty$ and hence sends $P \to 0$. We therefore expect the limiting solution to live in the $\mathcal P=0$ regime of the constant-pressure null sector discussed in Section~\ref{sec:nulllimitofrelfluids}. To verify this expectation directly from the bulk perspective, we compute the renormalised Brown--York tensor of the AdS-pp-wave metric~\eqref{eq:dsAdSnullideal} using~\eqref{eq:AdSTabdefn}. One finds
\begin{equation}
	\label{eq:nullTabAdSideal}
	\mathcal{T}_{ab}^{(0)}=\mathcal{E}v_{a}v_{b}\,, \qquad \mathcal{E}=\frac{d\kappa^{d}}{16\pi G_D}\,,
\end{equation}
which is precisely the energy-momentum tensor of a pressureless null fluid. In particular, the single ideal-order transport coefficient $\mathcal E$ is completely fixed by the bulk parameter $\kappa$, which is also the case in the usual fluid/gravity correspondence.

Moreover, the result makes manifest the ideal-order rescaling redundancy discussed in Section~\ref{sec:nullfluids}: the geometry, and hence $\mathcal{T}^{(0)}_{ab}$, depends only on the invariant combination $\kappa^{d}v_av_b$, so that a local rescaling of $v_a$ accompanied by a compensating rescaling of $\kappa$ leaves the ideal-order data unchanged. Finally, both the bulk construction of the AdS-pp-wave geometry (via $b\to\infty$) and the ultra-relativistic scaling limit of the parent timelike fluid imply that the limiting temperature $T_L$ is zero, ensuring that the gravitational and hydrodynamic limits are mutually consistent. We will see in Section~\ref{sec:landauframefromnulllimit} how the corresponding scaling extends to first order in derivatives.

\subsection{First-order metric}
\label{sec:AdSbottomup}

We now construct the first-derivative corrections to the ideal-order AdS-pp-wave geometry~\eqref{eq:dsAdSnullideal}, following the general logic of fluid/gravity~\cite{Bhattacharyya:2007vjd}. The ideal geometry is parametrised by a scale $\kappa$ and a null direction $v_a$. In the hydrodynamic regime we promote these collective variables to slowly varying fields on the boundary
\begin{equation}
	\{\kappa,v_a\} \longrightarrow \{\kappa(\sigma),v_a(\sigma)\}\,,
\end{equation}
and organise the expansion by a formal derivative-counting parameter $\epsilon$ such that
\begin{equation}
	\D_a \kappa \sim \mathcal{O}(\epsilon)\,,\qquad \text{and}\qquad \D_a v_b\sim \mathcal{O}(\epsilon)\,.
\end{equation}
Indices on boundary tensors are raised/lowered with the Minkowski metric $\eta_{ab}$ at this order. The derivative-counting parameter $\epsilon$ measures the ratio between the microscopic length scale set by $\kappa^{-1}$ and the macroscopic variation scale $\mathcal R$ of the fields 
\begin{equation}
	\epsilon \sim \frac{1}{\kappa\mathcal R}\ll 1\,.
\end{equation}
In a neighbourhood of a point $\sigma=0$ one may Taylor expand as 
\begin{equation}
	\label{eq:vkcollvar}
	\kappa(\sigma)=\kappa(0)+\epsilon\sigma^{a}\D_{a}\kappa(0)+\mathcal{O}(\epsilon^2)\,, \qquad v_{a}(\sigma)=v_{a}(0)+\epsilon\sigma^{b}\D_{b}v_{a}(0)+\mathcal{O}(\epsilon^2)\,,
\end{equation} and we will often choose coordinates adapted to the zeroth-order null direction so that $v_a(0)$ lies in the $(t,z)$-plane, i.e., $v_a(0)= (v_t(0),0_i,v_z(0))$. The null condition $v^av_a=0$ is imposed order by order: at ideal order, it implies that $v_z(0) = \pm v_t(0)$. Choosing the positive sign and $v_t(0)=1$, we may write $v_a = v_a(0) + \delta v_a$, where $\delta v_a$ is a generic first-order perturbation. The condition $v_a v^a = 0$ at first order implies that $v^a(0) \delta v_a= 0$, or, equivalently, $\delta v_t = \delta v_z$. This implies that the vector form of $v_a(\sigma)$ as defined in~\eqref{eq:vkcollvar} reads 
\begin{equation}
\label{eq:vector-form}
\begin{split}
       v_a(\sigma) &= v_a(0) + \epsilon \,v^{(1)}_a(\sigma)\,,\\ 
       v_a(0) &= (1,0_i,1)\,,\qquad v^{(1)}_a(\sigma) = (\sigma^{b}\D_{b}v_{t}(0),\sigma^{b}\D_{b}v_{i}(0),\sigma^{b}\D_{b}v_{t}(0))\,,
\end{split}
\end{equation} 
to first order in the derivative expansion.

\paragraph{Ansatz and decomposition of Einstein's equations.}
Because the ideal-order geometry is nonsingular and carries no finite-temperature horizon, we do not impose horizon regularity conditions. Instead we work in a convenient radial gauge and impose asymptotically locally AdS boundary conditions. We take the first-order corrected metric to be of the form
\begin{equation}
	\label{eq:dsAdSnullbottomup}
	ds^2=\frac{dr^{2}}{r^{2}}
	+r^{2}\left(\eta_{ab}+\frac{\kappa(\sigma)^{d}}{r^d}v_{a}(\sigma)v_{b}(\sigma)\right)d\sigma^a d\sigma^b
	+f_{\mu\nu}(r)\,dx^\mu dx^\nu\,,
\end{equation}
where $f_{\mu\nu}\sim\mathcal{O}(\epsilon)$ encodes the corrections required for the metric to solve Einstein's equations once $(\kappa,v_a)$ vary. At first order, it is consistent to take $f_{\mu\nu}$ to depend on $r$ only, with $\sigma$-dependence present only in first-derivative data multiplying the radial integration functions that arise when solving the ODEs.

Einstein's equations split into constraints and dynamical equations. The constraints are the mixed radial components $E_{ra}=0$, which at first derivative order reduce to
\begin{equation}
	\label{eq:constrainteqnsads}
	\D_a(\kappa^d v^a)=\tau_a\dot v^a\,,\qquad h^{ab}\dot{v}_a=0\,,
\end{equation}
where now
\begin{equation}
	\dot v_a:=v^b\D_b v_a\,,\qquad \vartheta:=\D_a v^a\,,
\end{equation} $\tau_a$ and $h_{ab}$ are the auxiliary null vector and orthogonal projector, respectively, that were defined in Section~\ref{sec:nullfluids}.\footnote{For the adapted coordinates considered here, these take the form $\tau_a=(1,0_i,-1)/2$ and $h_{ab}=\delta_{ij}$. Since $\dot v_a$ and $\vartheta$ are both $\mathcal{O}(\epsilon)$, we do not need to consider derivative corrections to $\tau_a$ and $h_{ab}$. }
As anticipated,~\eqref{eq:constrainteqnsads} reproduces the ideal-order null-fluid equations~\eqref{eq:eomESBP}.
The remaining components $E_{\mu\nu}=0$ provide dynamical equations that determine $f_{\mu\nu}$.

A useful simplification is that, at first derivative order, the radial constraint equations~\eqref{eq:constrainteqnsads}
eliminate the inhomogeneous source terms in the remaining components of Einstein's equations. Equivalently, once
\eqref{eq:constrainteqnsads} holds, the dynamical equations for $f_{\mu\nu}$ reduce to homogeneous radial ODEs.
It follows that choosing the trivial homogeneous solution $f_{\mu\nu}=0$ is consistent at $\mathcal O(\epsilon)$.
More generally, one may add the full set of homogeneous modes and then retain only those compatible with radial gauge
and asymptotically locally AdS boundary conditions. This yields the most general first-order solution space, which we now construct.

\paragraph{Solving the dynamical equations for $d\geq3$.} We now solve the Einstein equations, whose explicit components are collected in Appendix~\ref{app:explicitEFE}. The simplest equations arise from the off-diagonal spatial components $E_{ij}$ with $i\neq j$, \eqref{eq:E_ij}, which is solved for\footnote{Note that this solution is for $d>3$. For $d=2,3$ this equation is absent.}
\begin{equation}
\label{eq:solfijads}
	f_{ij}=r^{2}c^{(2)}_{ij} + \frac{c_{ij}^{(1)}}{r^{d-2}}\,,
\end{equation}
with integration functions $c^{(1,2)}_{ij}(\sigma)\sim\mathcal{O}(\epsilon)$.
Asymptotically locally AdS boundary conditions with fixed boundary metric require the non-normalisable $r^2$ mode to vanish, and hence $c^{(2)}_{ij}=0$. The components $E_{ti}$ and $E_{iz}$, \eqref{eq:E_ti-E_iz}, are coupled, and solving the resulting radial system yields 
\begin{equation}\label{eq:solftifziads}
	\begin{split}
		f_{ti}&=r^2 \left(-\frac{c^{(2)}_{iz}}{\kappa^d}+c^{(1)}_{ti}+c^{(3)}_{iz}\right)
		+\frac{c^{(2)}_{iz}-2 c^{(1)}_{iz}/\kappa^d}{r^{d-2}}+\frac{c^{(1)}_{iz}}{r^{2d-2}}\,,\\
		f_{iz}&=c^{(3)}_{iz} r^2+\frac{c^{(2)}_{iz}}{r^{d-2}}+\frac{c^{(1)}_{iz}}{r^{2d-2}}\,,
	\end{split}
\end{equation}
together with the identification
\begin{equation}
	\label{eq:cond1}
	c_{ti}^{(2)}=c_{iz}^{(2)}\,,
\end{equation}
which follows from Einstein's equations. Again, asymptotically locally AdS boundary conditions fix the coefficients multiplying $r^2$ to vanish, setting $c^{(3)}_{iz} = 0$  and $c^{(3)}_{ti} = \frac{c^{(2)}_{iz}}{\kappa^d}$.

The remaining components $E_{tt}$, $E_{tz}$, $E_{zz}$, $E_{ii}$ and $E_{rr}$ are coupled.
For $d>3$, the difference $E_{ii}-E_{jj}$ yields a simple decoupled equation which implies that the diagonal spatial components can be written as (no sum over $i$)
\begin{equation}
	\label{eq:solf_ii}
	f_{ii}=%
    \frac{c^{(1)}_{ii}}{r^{d-2}}+\chi(r)\,,
\end{equation}
where $\chi(r)$ is a residual function common to all diagonal components.
For $d=3$ we keep the parametrisation~\eqref{eq:solf_ii} as well, with $\chi(r)$ included so as not to restrict generality. Solving the remaining coupled system determines $f_{tt}$, $f_{tz}$, $f_{zz}$, $f_{rr}$ and $\chi$ in terms of integration functions.
The explicit intermediate equations are lengthy (an example for $d=4$ is given in \eqref{eq:EttEtz}), but the solutions with nonrenormalisable $r^2$-modes that violate boundary conditions removed can be written compactly for all $d\geq 3$ as
\begin{equation}
\label{eq:solttzzrrads}
	\begin{split}
		f_{tt}&=
        \frac{ c^{(1)}_{tt}+2 c^{(1)}_{tz}+ c^{(1)}_{zz}}{ r^{d-2}}
		-\frac{(d-2)\kappa^d \chi}{2r^d}-\frac{\kappa^d\left(c^{(1)}_{tt} + \delta^{ij}c^{(1)}_{ij}\right)}{2 r^{2d-2}} 
		+\frac{c^{(1)}_{tt} \kappa^{2d}}{c_d r^{3d-2}}-\chi\,,\\
		f_{tz}&=
        \frac{c^{(1)}_{tz}+c^{(1)}_{zz}}{r^{d-2}}
		-\frac{(d-2)\kappa^d \chi}{2r^d}-\frac{\kappa^d \delta^{ij}c^{(1)}_{ij}}{2 r^{2d-2}}
		+\frac{c^{(1)}_{tt} \kappa^{2d}}{c_d r^{3d-2}}\,,\\
		f_{zz}&=
        \frac{c^{(1)}_{zz}}{r^{d-2}}
		-\frac{(d-2)\kappa^d \chi}{2r^d}+\frac{\kappa^d\left(c^{(1)}_{tt} - \delta^{ij}c^{(1)}_{ij} \right)}{2 r^{2d-2}}
		+\frac{c^{(1)}_{tt} \kappa^{2d}}{c_d r^{3d-2}}+\chi\,,\\
		f_{rr}&=\frac{\chi'}{r^3}-\frac{2 \chi}{r^4}
		+\frac{ c^{(2)}_{tt} \kappa^d+ (c^{(1)}_{tt}+2 c^{(1)}_{tz}-\delta^{ij}c^{(1)}_{ij})}{ r^{d+2}}
		-\frac{dc^{(1)}_{tt} \kappa^d}{(2d-2) r^{2d+2}}\,,
	\end{split}
\end{equation}
where
\begin{equation}
\label{eq:cd}
    c_d=\frac{12(d-1)}{(d-2)}\,,
\end{equation}
is a dimension-dependent parameter.\footnote{Note that the expression for $c_d$ in~\eqref{eq:cd} was only explicitly checked up to $d=5$. It would be interesting to confirm whether it remains true for $d\geq6$.} The function $\chi$, which satisfies $\chi(r)/r^2 \to 0$ as $r\to\infty$ to preserve boundary conditions, will not contribute to the boundary energy-momentum tensor (see below), so we may set it to zero without loss of generality. Finally, the components $f_{ra}$ are pure gauge in our setup and we work in radial gauge, setting
\begin{equation}
\label{eq:gauge-choice}
	f_{ra}=0\,.
\end{equation}
We discuss this in more detail in Appendix~\ref{app:fraremoval}, where we also demonstrate that the function $\chi(r)$ corresponds to a residual coordinate transformation that preserves the gauge choice, and that fixing $\chi(r)$ simply corresponds to fixing that residual transformation.

\paragraph{Boundary stress tensor and matching to null hydrodynamics.}
We now determine which choices of the remaining integration functions correspond to physical hydrodynamic perturbations of a null fluid.
The boundary energy-momentum tensor is obtained from the renormalised Brown--York prescription~\eqref{eq:AdSTabdefn}. It is convenient to isolate the first-order correction by subtracting the ideal contribution and rescaling by $16\pi G_D$
\begin{equation}
	\tilde{\mathcal{T}}_{ab}:=16\pi G_D\big(\mathcal T_{ab}-\mathcal T^{(0)}_{ab}\big)=16\pi G_D \mathcal T^{(1)}_{ab}\,, 	\qquad 	16\pi G_D\mathcal T^{(0)}_{ab}=d\kappa^d v_a v_b\,.
\end{equation}
In terms of the radial integration functions $c^{(\cdot)}(\sigma)$, the components of $\tilde{\mathcal T}_{ab}$ are
\begin{equation}
	\label{eq:stresscomponentsads}
	\begin{split}
		\tilde{\mathcal{T}}_{tt} &= -(d-1)c_{tt}^{(1)}-(2d-2)c_{tz}^{(1)}-\delta^{ij}c^{(1)}_{ij}-dc_{zz}^{(1)}
        \,, \\
		\tilde{\mathcal{T}}_{tz}&= -d\left(c^{(1)}_{tz}+c^{(1)}_{zz}\right)\,, \\
		\tilde{\mathcal{T}}_{ti} &= -d\left(c^{(2)}_{iz}-\frac{2 c^{(1)}_{iz}}{\kappa^d}\right)\,,\\
		\tilde{\mathcal{T}}_{zi} &= -dc^{(1)}_{iz}\,, \\
		\tilde{\mathcal{T}}_{ij}&=-dc^{(1)}_{ij}-\delta_{ij}\left(c_{tt}^{(1)}+2c_{tz}^{(1)}-\delta^{kl}c^{(1)}_{kl}
        \right)\, .
\end{split}\end{equation}
Importantly, $\tilde{\mathcal{T}}_{ab}$ is independent of $\chi$, confirming that $\chi$ is physically irrelevant for the boundary data. For simplicity, we shall henceforth set $\chi=0$. 

To match the bulk integration functions to null hydrodynamic data, we now express the radial integration ``constants''
$c^{(\cdot)}(\sigma)$ in terms of first-derivative structures built from the collective fields $(\kappa(\sigma),v_a(\sigma))$. Since our intermediate solution of the radial ODEs was obtained in a local patch adapted to the equilibrium null direction, we temporarily work in boundary coordinates $\sigma^a=(t,x^i,z)$ chosen such that at $\epsilon=0$ one has $v_a(0)=(1,0,\ldots,0,1)$, i.e., the null direction lies in the $(t,z)$-plane.
In such an adapted patch it is natural to allow the full set of first-order component derivatives
$\partial_a\kappa$ and $\partial_a v_b$ as a basis, with covariance only emerging after the coefficients are fixed.

At first order, the independent null fluid data may be organised into scalars, vectors and tensors according to
\begin{equation}
	\label{eq:firstorderdata}
	\begin{split}
		&\text{scalars:}\ \vartheta\,,\ v^a\partial_a\kappa\,,\ (c^{(1)}_{tt},\ c^{(1)}_{zz}\,,\ c^{(1)}_{tz}
        )\, , \\
		&\text{vectors:}\ \dot v_a\,,\ \partial_a\kappa\,,\ (c^{(1)}_{iz}\,,\ c^{(2)}_{ti} 
        )\,,\\ 
		&\text{tensors:}\ \varsigma_{ab}\,, v_{(a}\partial_{b)}\kappa\,,\ (c^{(1)}_{ij})\,. 
	\end{split}
\end{equation}
In particular, in the scalar sector we may take $\vartheta=\D_a v^a$ and $v^a\D_a\kappa$ as the covariant scalars,
but in an adapted patch it is also convenient to keep track of the separate components $\partial_t\kappa$ and $\partial_z\kappa$.
Accordingly, we parametrise the scalar integration functions as
\begin{equation}
	\label{eq:scalar-ansatz-aN}
	\begin{split}
		c^{(1)}_{tt}&=a_{1}\vartheta+a_2\frac{v^a\D_a\kappa}{\kappa}
		+\frac{2(a_3\partial_t+a_4\partial_z)\kappa}{\kappa}\,, \\
		c^{(1)}_{zz}&=a_{5}\vartheta+a_6\frac{v^a\D_a\kappa}{\kappa}
		+\frac{2(a_7\partial_t+a_8\partial_z)\kappa}{\kappa}\,,\\
		c^{(1)}_{tz}&=a_{9}\vartheta+a_{10}\frac{v^a\D_a\kappa}{\kappa}
		+\frac{2(a_{11}\partial_t+a_{12}\partial_z)\kappa}{\kappa}\,, 
	\end{split}
\end{equation}
where the coefficients $a_\bullet(\kappa)$ are functions of $\kappa$. The terms involving $\partial_t\kappa$ and $\partial_z\kappa$ are non-covariant artefacts of working in this adapted patch, and are included here only to maintain a general component basis at this intermediate stage. Requiring that the final answer be covariant for both the metric correction and the boundary energy-momentum tensor will force the corresponding
coefficients to vanish, or, equivalently, to recombine into the covariant scalar $v^a\D_a\kappa$. In the vector sector, the coupled solution~\eqref{eq:solftifziads} together with the condition~\eqref{eq:cond1} restricts the allowed combinations. A convenient parametrisation obeying~\eqref{eq:cond1} is
\begin{equation}
	\label{eq:vector-ansatz-aN}
	\begin{split}
		c^{(2)}_{ti}&=-2\Big(a_{13}v_{(t}\dot v_{i)}+a_{14}\frac{v_{(t}\D_{i)}\kappa}{\kappa}
		+a_{15}\partial_{(z}v_{i)}\Big)\, , \\
		c^{(1)}_{iz}&=\kappa^{d}a_{15}\dot v_i\,,
	\end{split}
\end{equation}
where, as alluded to above, $\partial_z v_i$ is retained as a bookkeeping device; in the final covariant form it will be absorbed into $\D_{(a}v_{b)}$. Finally, in the symmetric tensor sector we have
\begin{equation}
	\label{eq:tensor-ansatz-aN}
	c^{(1)}_{ij}=-2\left(a_{16}\partial_{(i}v_{j)}+a_{17}\frac{v_{(i}\D_{j)}\kappa}{\kappa}+a_{18}v_{(i}v_{j)}\vartheta\right)\,.
\end{equation}
Altogether, this parametrisation introduces $18$ a priori independent coefficients $a_\bullet(\kappa)$.
These will be fixed by requiring (i) asymptotically locally AdS boundary conditions with fixed boundary metric,
(ii) a finite renormalised Brown--York stress tensor, and (iii) that both $f_{\mu\nu}$ and $\mathcal T_{ab}$ can be written
covariantly in terms of first-order null-fluid data.
These requirements lead to relations among the $a_\bullet$ and eliminate the non-covariant patch artefacts,
leading to a parametrisation in terms of dimensionless constants $\tilde a_\bullet$ defined via 
\begin{equation}
	\label{eq:afixads}
	\begin{split}
		a_{1,2,3,4,7,9,11,12}&=0\, , ~~
		a_{15}=a_{16}=2a_{18}=-\kappa^{d+1}\frac{\tilde a_\eta}{d}=\kappa^{d+1}\frac{\tilde a_1}{2} \, ,\\
		a_{5}&=-2\kappa^{d+1}\frac{\tilde a_2}{d}\, , ~~
		a_{8}=-a_{10}=
        a_{14}=a_{17}=-\kappa^{d+1}\frac{\tilde a_5}{d}\, , \\
		a_{6}&=-2\kappa^{d+1}\frac{\tilde a_4}{d}\, ,~~
		a_{13}=-\kappa^{d+1}\frac{\tilde a_3}{d}\, ,~~
		\tilde a_6=-\frac{2}{d}\tilde a_5\, .
	\end{split}
\end{equation}

\paragraph{Resulting first-order geometry and energy-momentum tensor.}
Collecting our results,\footnote{The results presented here have been explicitly checked for $d\leq5$.} the first-order asymptotically AdS null metric may be written as~\eqref{eq:dsAdSnullbottomup} with correction
\begin{equation}
	\label{eq:mostgenfmunuads}
	\begin{split}
		f_{ab}&=-\frac{2\kappa^{d+1}}{dr^{d-2}}\left(\left(\tilde a_2\vartheta +\tilde a_4 \frac{v^c\D_c\kappa}{\kappa}\right)v_av_b+\tilde a_3v_{(a}\dot v_{b)}-\tilde a_6\frac{dv_{(a}\D_{b)}\kappa}{2\kappa}+\tilde a_\eta\D_{(a}v_{b)}\right)\\
		&\quad+\frac{\kappa^{2d+1}}{dr^{2d-2}}\tilde a_\eta\left(\vartheta v_a v_b-v_{(a}\dot v_{b)} \right)\, , \\
		f_{rr}&=\frac{2\kappa^{d+1}}{dr^{d+2}}\left(\tilde a_\eta\vartheta -\tilde a_6 \frac{dv^a\D_a\kappa}{2\kappa}\right)\, .
	\end{split}
\end{equation}
We have checked explicitly that the Kretschmann scalar remains unchanged at this order, $	K_r=2d(d+1)+\mathcal{O}(\epsilon^2)$, so the spacetime remains locally AdS-like and has no horizon at first derivative order. Moreover, we have explicitly checked that the solutions above are also valid for $d=2$. The associated boundary energy-momentum tensor (for $d\geq2$) takes the form $\mathcal{T}^{ab}=\mathcal{T}^{ab}_{(0)}+\mathcal{T}^{ab}_{(1)}$, with $\mathcal{T}^{ab}_{(0)}$ given in~\eqref{eq:nullTabAdSideal} and
\begin{equation}
	\label{eq:mostgenstressads}
	\mathcal{T}^{ab}_{(1)}=\left(\rho_1\vartheta+\rho_6\frac{v^c\D_c\kappa}{\kappa}\right) \eta^{ab}
	+2v^{(a}\left(\rho_2\vartheta v^{b)}+\rho_3\dot{v}^{b)}+\rho_4v^{b)}\frac{v^c\D_c\kappa}{\kappa}+\rho_5\frac{\D^{b)}\kappa}{\kappa}\right)-2\eta\varsigma^{ab}\,,
\end{equation}
where
\begin{equation}
	\rho_\bullet=-\frac{\kappa^{d+1}\tilde a_\bullet}{16\pi G_D}\, ,\qquad
	\eta=\frac{\kappa^{d+1}\tilde a_\eta}{16\pi G_D}\, ,\qquad
	\rho_1=\frac{2}{d}\eta\, ,\qquad
	\rho_6=-\frac{2}{d}\rho_5\, .
\end{equation}
This matches~\eqref{eq:firstnullTgen} subject to the gravitational constraints among transport coefficients, in close analogy with the timelike fluid/gravity case~\cite{Bhattacharyya:2007vjd,Haack:2008cp}. In particular, these constraints ensure that the stress tensor is traceless, $\mathcal{T}^a{_a}=0$, as expected for a conformal theory on the AdS boundary. Moreover, evaluating $\D_a\mathcal{T}^{ab}=0$ reproduces the constraint equations~\eqref{eq:constrainteqnsads}, confirming that the bulk constraints are precisely the null hydrodynamic equations.

\paragraph{Bulk frame transformations.}
As argued in Section~\ref{sec:nullfluids}, frame transformations can be used to remove redundant first-order structures and reduce the number of independent transport coefficients for the null fluid. The bulk dual admits an analogous transformation (see Appendix~\ref{app:frameandcoord}), where an on-shell first-order field redefinition of $(\kappa,v^a)\to (\kappa + \delta \kappa,v^a + \delta v^a)$ induces a shift in $f_{\mu\nu}$. By choosing $(\delta\kappa,\delta v^a)$ appropriately, we can rewrite the same bulk metric in such a way that certain first-derivative structures are absent. Concretely, choosing the on-shell transformation
\begin{equation}
\label{eq:frametonullL}
	\delta g_{ab}(\delta v,\delta\kappa)=\delta g_{ab}\left(\tilde a_3\frac{\kappa}{d}\dot{v},\frac{2\kappa^2(\tilde a_2 \vartheta+\tilde a_4 v^b\D_b\kappa/\kappa)}{d^2}\right)\,,
\end{equation}
and using~\eqref{eq:pertmetric}, sets $\tilde a_{2,3,4}=0$ in~\eqref{eq:mostgenfmunuads}, which correspondingly eliminates $\rho_{2,3,4}$ from~\eqref{eq:mostgenstressads}. In this frame the gravitational calculation fixes the remaining transport coefficients uniquely.

\subsection{Specific frames from null limit}
\label{sec:landauframefromnulllimit}

In this subsection we show that the asymptotically AdS solutions constructed bottom-up in Section~\ref{sec:AdSbottomup} are reproduced by taking a null limit of the standard first-order fluid/gravity metrics dual to boosted AdS black branes. The required timelike first-order solutions were obtained in~\cite{Bhattacharyya:2007vjd,Haack:2008cp} in Eddington--Finkelstein-like (EF-like) coordinates and in Landau frame.

EF-like coordinates are natural in the timelike fluid/gravity construction because they make horizon regularity manifest. In the case of geometries dual to null fluids, however, the limiting geometry has no horizon, making it convenient to work in an AdS--Schwarzschild-like/radial gauge
adapted to the pp-wave form~\eqref{eq:dsAdSnullideal} and to the gauge choices used in Section~\ref{sec:AdSbottomup}. 

Starting from the first-order fluid/gravity metrics of~\cite{Bhattacharyya:2007vjd,Haack:2008cp}, we first rewrite the perturbed boosted black brane in a Schwarzschild-like (radial) coordinate system adapted to our null limit and to the gauge choices used in Section~\ref{sec:AdSbottomup}. Concretely, one may obtain this form either by an explicit coordinate transformation from EF-like coordinates or, more efficiently, by using the AdS/Ricci-flat correspondence of~\cite{Caldarelli:2013aaa} to map to a convenient representation and then mapping back. The details of this rewriting, and the resulting Schwarzschild-like expression we use in the limit, are given in Appendix~\ref{app:firstads}.

\paragraph{Double scaling in the null limit.}
At ideal order the null limit is implemented by sending $b\to\infty$ (cf.~Section~\ref{sec:idealads}), and using~\eqref{eq:TsAdS} this corresponds to the low-temperature limit $T\sim b^{-1}\to 0$.
The timelike derivative expansion is controlled by $\mathcal{R}T\gg 1$, where $\mathcal{R}$ is the macroscopic variation scale of the hydrodynamic fields. Thus, to retain a controlled hydrodynamic regime while sending $b\to\infty$, we must simultaneously take $\mathcal{R}$ large, and  we parametrise this by assuming a double scaling where $\mathcal{R}$ scales as a power of $b$ such that $\mathcal{R}T\gg 1$. 

This is the gravitational counterpart of the discussion in Section~\ref{sec:nulllimitofrelfluids}: the null limit must be taken together with a scaling of gradients. In the following, we shall take the limit in two different timelike frames. As discussed in Section~\ref{sec:nulllimitofrelfluids}, the resulting null fluid data, and hence the corresponding dual bulk metric, depends on this choice of frame. 

\subsubsection*{Landau frame} 
The first-order timelike fluid/gravity metrics of~\cite{Bhattacharyya:2007vjd,Haack:2008cp} are written in Landau frame. Taking the null limit directly in this frame therefore produces only a particular subfamily of our general first-order solution~\eqref{eq:mostgenfmunuads}. From the hydrodynamic analysis in Section~\ref{sec:nulllimitofrelfluids} (cf.~\eqref{eq:limitT1landau}),
the limiting stress tensor is of the form~\eqref{eq:mostgenstressads} with
\begin{equation}
	\rho_{1,4,5,6}=0\,,\qquad \eta=0\,,
\end{equation}
so that only the $(\rho_2,\rho_3)$ structures survive. Equivalently, on the gravity side this corresponds to setting
$\tilde a_{1,4,5,6,\eta}=0$ in~\eqref{eq:mostgenfmunuads}, i.e.,
\begin{equation}
	\label{eq:landaulimfmunuads}
	f_{ab}=-\frac{2\kappa^{d+1}}{dr^{d-2}}\left(\tilde a_2\,\vartheta\, v_a v_b+\tilde a_3\, v_{(a}\dot v_{b)}\right)\,, \qquad f_{rr}=0\,.
\end{equation}
We now demonstrate that the null limit of the timelike perturbed black brane indeed reproduces~\eqref{eq:landaulimfmunuads} and fixes the pair $(\tilde a_2,\tilde a_3)$.

\paragraph{Hydrodynamic scaling.}
Using the ideal-order relation $\gamma\sim \mathcal{O}(b^{d/2})$, one may evaluate the scaling limit~\eqref{eq:limitT1landau} explicitly from the timelike first-order energy-momentum tensor~\eqref{eq:Tab1AdSnonnull}, leading to
\begin{equation}
	\label{eq:speclimlandauads}
	\begin{split}
		2\hat\eta\,\gamma^{\alpha+3}\,\frac{U_aU_b\,\partial_cU^c}{d-1}
		&\sim
		\frac{b^{\frac{d}{2}(\alpha+3)-(d-1)}}{8\pi G_D}\,\frac{U_aU_b\,\partial_cU^c}{d-1}
		\rightsquigarrow
		\frac{\kappa^{d+1}}{8(d-1)\pi G_D}\,v_av_b\,\vartheta
		=: \rho_{2}\,v_av_b\,\vartheta\,,
		\\
		2\hat\eta\,\gamma^{\alpha+3}\,U_{(a}\dot U_{b)}
		&\sim
		\frac{b^{\frac{d}{2}(\alpha+3)-(d-1)}}{8\pi G_D}\,U_{(a}\dot U_{b)}
		\rightsquigarrow
		\frac{\kappa^{d+1}}{8\pi G_D}\,v_{(a}\dot v_{b)}
		=: 2\rho_{3}\,v_{(a}\dot v_{b)}\,,
	\end{split}
\end{equation}
which is finite and non-trivial for
\begin{equation}
	\alpha=-\frac{d+2}{d}\,.
\end{equation}
Note that the overall power $\kappa^{d+1}$ is fixed by dimensional analysis and agrees with the gravitational normalisation of~\eqref{eq:mostgenstressads}.

\paragraph{Gravitational limit.}
On the bulk side we start from the timelike perturbed black brane metric in Schwarzschild-like form~\eqref{eq:transfAdSmetric}.
To compare with the radial gauge used in Section~\ref{sec:AdSbottomup} and to obtain a smooth null limit, one must eliminate the mixed components $f_{ra}$
by a (first-order) coordinate transformation; the explicit procedure is described in Appendix~\ref{app:fraremoval}.
One then takes the null limit with $\gamma\sim b^{d/2}$ and the same value $\alpha=-(d+2)/d$ as in~\eqref{eq:speclimlandauads}.
For example, the term proportional to $f_1$ in~\eqref{eq:fmununonnullads} behaves as
\begin{equation}
	\label{eq:limfabadslandau}
	2f_{1} u_a u_b \theta
	\sim
	2f_1 b^{\frac{d}{2}(\alpha+3)} U_aU_b \partial_cU^{c}
	\rightsquigarrow
	-\frac{2\kappa^{d+1}}{d(d-1)\,r^{d-2}} v_av_b \vartheta\,,
\end{equation}
and similarly for the remaining first-order structures. Carrying this out for all components yields
\begin{equation}
	f_{ab}=\frac{2\kappa^{d+1}}{dr^{d-2}}
	\left(\frac{1}{d-1} \vartheta  v_a v_b+ v_{(a}\dot v_{b)}\right)\,, \qquad	f_{rr}=0\,,
\end{equation}
which matches~\eqref{eq:landaulimfmunuads} provided
\begin{equation}
	\tilde a_2=-\frac{1}{d-1}\,,\qquad \tilde a_3=-1\,.
\end{equation}
Equivalently, the limiting boundary energy-momentum tensor is precisely~\eqref{eq:mostgenstressads} with
$\rho_{1,4,5,6}=\eta=0$ and $(\rho_2,\rho_3)$ as in~\eqref{eq:speclimlandauads}, as required.
Finally, acting with the null-frame transformation~\eqref{eq:frametonullL} eliminates the surviving $(\rho_2,\rho_3)$ structures,
yielding a representative metric with $f_{\mu\nu}=0$. This mirrors the frame redefinition freedom discussed for null fluids in Section~\ref{sec:nullfluids}.

\subsubsection*{Non-thermodynamic frame}

We next take the null limit after first performing a timelike frame transformation away from the Landau frame, as described in Section~\ref{sec:nulllimitofrelfluids}.
At the level of the timelike variables, the required redefinition can be implemented by
\begin{equation}
	\label{eq:hydroframeads}
	\delta u_a = \frac{2b\dot{u}_a-\partial_ab-u_au^c\partial_cb}{2d}\,,
	\qquad
	\delta b= \frac{2b^2\theta-(d-1)b\,u^c\partial_cb}{d^2(d-1)}\,,
\end{equation}
so that the first-order energy-momentum tensor takes the non-thermodynamic form~\eqref{eq:newnonthermoT}, namely
\begin{equation}
	T_{ab}=T_{ab}^{(0)}-2\hat\eta\left(\partial_{(a}u_{b)}+\frac{u_{(a}\partial_{b)}b}{b}
	+\left(\frac{u^c\partial_cb}{2db}-\frac{\theta}{d}\right)\eta_{ab}\right)\,.
\end{equation}

\paragraph{Hydrodynamic scaling.} 
Taking the null limit now keeps fixed the combinations appropriate to $h=0$ structures (cf.~\eqref{eq:limitT1nulllandu}).
Using again $\gamma\sim \mathcal{O}(b^{d/2})$, one obtains
\begin{equation}
	\label{eq:speclimnonlandauads}
	\begin{split}
		2\hat\eta\,\gamma^{\alpha+1}\,\partial_{(a}U_{b)}
		&\sim
		\frac{b^{\frac{d}{2}(\alpha+1)-(d-1)}}{8\pi G_D}\,\partial_{(a}U_{b)}
		\rightsquigarrow
		\frac{\kappa^{d+1}}{8\pi G_D}\,\partial_{(a}v_{b)}
		=: 2\eta\,\varsigma_{ab}\,,
		\\
		\frac{2}{d}\hat\eta\,\gamma^{\alpha+1}\,\partial_{a} U^{a}
		&\sim
		\frac{b^{\frac{d}{2}(\alpha+1)-(d-1)}}{8d\pi G_D}\,\partial_{a} U^{a}
		\rightsquigarrow
		\frac{\kappa^{d+1}}{8d\pi G_D}\,\vartheta
		=: 2\rho_{1}\,\vartheta\,,
		\\
		\frac{2}{ b}\,\hat\eta\,\gamma^{\alpha+1}\,U_{(a}\partial_{b)}b
		&\sim
		\frac{b^{\frac{d}{2}(\alpha+1)-(d-1)}}{8\pi G_D}\,U_{(a}\partial_{b)}\kappa
		\rightsquigarrow
		\frac{\kappa^{d+1}}{8\pi G_D}\,\frac{v_{(a}\partial_{b)}\kappa}{\kappa}
		=: 2\rho_5\,\frac{v_{(a}\partial_{b)}\kappa}{\kappa}\,,
		\\
		\frac{1}{db}\,\hat\eta\,\gamma^{\alpha+1}\,U^{a}\partial_{a}b
		&\sim
		\frac{b^{\frac{d}{2}(\alpha+1)-(d-1)}}{16d\pi G_D}\,U^{a}\partial_{a}\kappa
		\rightsquigarrow
		\frac{\kappa^{d+1}}{16d\pi G_D}\,\frac{v^{a}\partial_{a}\kappa}{\kappa}
		=: \rho_6\,\frac{v^{a}\partial_{a}\kappa}{\kappa}\,,
	\end{split}
\end{equation}
which is finite and non-trivial for
\begin{equation}
	\alpha=\frac{d-2}{d}\,.
\end{equation}
The limiting energy-momentum tensor is therefore~\eqref{eq:mostgenstressads} with $\rho_{2,3,4}=0$, as expected for the non-thermodynamic frame.

On the gravity side this corresponds to setting $\tilde a_{2,3,4}=0$ in~\eqref{eq:mostgenfmunuads}, i.e.,
\begin{equation}
	\label{eq:fabnonthermoads}
	\begin{split}
		f_{ab}&=-\frac{2\kappa^{d+1}}{dr^{d-2}}
		\left(-\tilde a_6\,\frac{d v_{(a}\partial_{b)}\kappa}{2\kappa}+\tilde a_\eta\,\partial_{(a}v_{b)}\right)
		+\frac{\kappa^{2d+1}}{dr^{2d-2}}\tilde a_\eta\left(\vartheta v_a v_b-v_{(a}\dot v_{b)} \right)\,,
		\\
		f_{rr}&=\frac{2\kappa^{d+1}}{dr^{d+2}}
		\left(\tilde a_\eta\,\vartheta -\tilde a_6\,\frac{d v^a\partial_a\kappa}{2\kappa}\right)\,.
	\end{split}
\end{equation}

\paragraph{Gravitational limit.}
To realise this limit starting from the timelike perturbed black brane, we first perform the timelike frame transformation described in Appendix~\ref{app:frames}, and then fix radial gauge by eliminating $f_{ra}$ as in Appendix~\ref{app:fraremoval}.
The resulting metric is of the form~\eqref{eq:dsSchnonnullfirst}, with shifted coefficients
\begin{equation}
	\begin{split}
		\tilde f_{ab}&=f_{ab}+\frac{1}{r^{d-2}b^{d}}\left(\frac{2b\,\dot u_{(a}-\partial_{(a}b}{d} u_{b)}
		-\frac{2b\theta}{d(d-1)} u_a u_b\right)\,,
		\\
		\tilde f_{rr}&=f_{rr}-\frac{2b\theta-(d-1)u^c\partial_cb}{d(d-1) r^{d+2}b^{d} f(br)^2}\,,
		\qquad
		\tilde f_{ra}=0\,,
	\end{split}
\end{equation}
where $f_{ab}$ and $f_{rr}$ are those of~\eqref{eq:fmununonnullads}. As an illustrative example, the term proportional to $\tilde f_1$ (with $\tilde f_1=f_1-1/(d(d-1)r^{d-2}b^{d-1})$) behaves as
\begin{equation}
	\label{eq:limfabadsnonlandau}
	2\tilde f_1\,u_au_b\,\theta
	\sim
	2\tilde f_1\,b^{\frac{d}{2}(\alpha+3)} U_{a}U_{b}\,\partial_cU^{c}
	\rightsquigarrow
	\frac{2\kappa^{2d+1}}{d\,r^{2d-2}} v_av_b \vartheta\,,
\end{equation}
which is finite precisely for the same value $\alpha=(d-2)/d$ found in~\eqref{eq:speclimnonlandauads}.
Proceeding analogously for the remaining terms gives
\begin{equation}
	\begin{split}
		f_{ab}&=-\frac{2\kappa^{d+1}}{dr^{d-2}}\left(\partial_{(a}v_{b)}+\frac{v_{(a}\partial_{b)}\kappa}{\kappa}\right)
		+\frac{\kappa^{2d+1}}{dr^{2d-2}}\left(\vartheta v_a v_b-v_{(a}\dot v_{b)} \right)\,,
		\\
		f_{rr}&=\frac{2\kappa^{d+1}}{dr^{d+2}}\left(\vartheta+\frac{v^c\partial_c\kappa}{\kappa} \right)\,,
	\end{split}
\end{equation}
which matches~\eqref{eq:fabnonthermoads} provided
\begin{equation}\label{eq:non-thermofix}
	\tilde a_\eta=1\,,\qquad \tilde a_6=-\frac{d}{2}\,.
\end{equation}
Thus, the null limit of the timelike perturbed black brane in this frame reproduces the non-thermodynamic metric and yields
$\mathcal{T}^{ab}_{(1)}$ of the form~\eqref{eq:mostgenstressads} with $\rho_{2,3,4}=0$ and $(\rho_1,\eta,\rho_5,\rho_6)$ as in~\eqref{eq:speclimnonlandauads}. In particular, as a consistency check, the double-scaling exponent $\alpha$ required for a finite, non-trivial null limit of the first-derivative data is the same in the hydrodynamic scaling analysis and in the bulk metric limit, corroborating that the two constructions describe the same theory of null hydrodynamics.

\section{Asymptotically flat null solutions}
\label{sec:asyflatnullsol}

In this section we perturbatively construct inhomogeneous asymptotically flat pp-wave solutions of the Einstein equations, which can be thought of as the Ricci-flat analogues of the AdS solutions we considered in Section~\ref{sec:asyadsnullsol}. We show that the Brown--York stress tensor associated with the resultant metrics is that of a null fluid discussed in Section~\ref{sec:nullhydrodynamics}. As in the AdS case, we show that particular types of these pp-wave solutions can be obtained by taking an ultra-relativistic limit of blackfolds, that is inhomogeneous asymptotically flat black branes.

\subsection{Ideal order from null limit}
\label{sec:idealflat}

We consider pure Einstein gravity in $D=d+1=p+n+3$ dimensions, where the brane worldvulime is $(p+1)$-dimensional, while the space transverse to the worldvolume is $(n+2)$-dimensional. At ideal order the relevant geometry is the neutral Schwarzschild black $p$-brane, boosted along the $(p+1)$-dimensional worldvolume. In Schwarzschild-like coordinates it can be written as~\cite{Emparan:2008eg}
\begin{equation}
	\label{eq:dsSch}
	ds^2=\left(\eta_{ab}+\frac{r_0^n}{r^n}u_{a}u_{b}\right)d\sigma^a d\sigma^b
	+\left(1-\frac{r_0^n}{r^n}\right)^{-1}dr^2+r^2d\Omega_{n+1}^2\, ,
\end{equation}
where $a=0,\dots,p$ labels worldvolume directions, and the constant worldvolume velocity satisfies $u^a u^b \eta_{ab}=-1$. We denote the horizon radius by $r_0$, and $d\Omega_{n+1}^2$ is the unit metric on the transverse $(n+1)$-sphere. The results below have been explicitly checked for $p=1,2,3,4$ and $n=1,2,3,4$ and therefore we expect them to hold for any $(p,n)$. 

\paragraph{Quasilocal energy-momentum tensor at infinity.}
The geometry~\eqref{eq:dsSch} carries conserved energy-momentum densities that can be extracted at spatial infinity, or, equivalently via ADM, or via a Brown--York subtraction prescription~\cite{Brown:1992br,Myers:1999psa,Camps:2010br}. For a timelike hypersurface at large radius $r=R$ with induced metric $\gamma_{\mu\nu}$, the subtracted Brown--York tensor is
\begin{equation}
	\label{eq:BYflat}
	T_{\mu\nu}^{\text{(BY)}}=\frac{1}{8\pi G_D}\Big[(K_{\mu\nu}-K\gamma_{\mu\nu})-(\bar K_{\mu\nu}-\bar K\gamma_{\mu\nu})\Big]\, ,
\end{equation}
where $K_{\mu\nu}$ is the extrinsic curvature of the $r=R$ surface embedded in~\eqref{eq:dsSch}, and $\bar K_{\mu\nu}$ is the corresponding quantity for the same induced geometry embedded in flat spacetime. Integrating over the transverse $\mathbb S^{n+1}$ and taking $R\to\infty$ yields an effective worldvolume energy-momentum tensor of perfect-fluid form
\begin{equation}
	\label{eq:TabdsSch}
	T_{ab}^{(0)}=(\varepsilon+P)u_{a}u_{b}+P\eta_{ab}\ , \qquad
	\varepsilon=-(n+1)P=(n+1)\frac{\Omega_{n+1}r_0^n}{16\pi G_D}\,.
\end{equation}
The solution also has local temperature and entropy density~\cite{Emparan:2009at}
\begin{equation}
	\label{eq:Tr0dsSch}
	T=\frac{n}{4\pi r_0}\, ,\qquad s=\frac{\Omega_{n+1}r_0^{n+1}}{4G_D}\, ,
\end{equation}
so one may equivalently parametrise the local thermodynamics by $r_0$ or $T$. In this way, the horizon radius $r_0$ is the flat-space analogue of the length scale $b$ we used in the AdS construction discussion in Section~\ref{sec:asyadsnullsol}. 

\paragraph{Aichelburg--Sexl-type ultraboost and null limit.}
We now take the ideal-order null limit as an ultra-relativistic boost along the worldvolume, accompanied by a vanishing-horizon-radius limit, $r_0 \to 0$. Writing $u_a=\gamma U_a$ with $U^aU^b\eta_{ab}=-1$ and $\gamma\to\infty$, we send
\begin{equation}
	\label{eq:r0LlimitdsSch}
	\gamma^{2}r_{0}^{n}\rightsquigarrow \kappa^n\,,\qquad
	U_a \rightsquigarrow v_a\,,\qquad \text{where}\qquad  v^a v^b\eta_{ab}=0\,,
\end{equation}
keeping $\kappa$ fixed, and where ``$\rightsquigarrow$'' denotes the double-scaling limit $(\gamma,r_0)\to (\infty,0)$. This is the natural generalisation of the Aichelburg--Sexl ultraboost~\cite{Aichelburg:1970dh} to the present brane set-up. Since $r_0$ has dimensions of length, so does $\kappa$, and the limit therefore implies that $T\to\infty$ through~\eqref{eq:Tr0dsSch}: in contrast to AdS, the null limit in flat space corresponds to an infinite-temperature limit. In this limit, the metric~\eqref{eq:dsSch} reduces to the asymptotically flat pp-wave
\begin{equation}
	\label{eq:dsSchnullideal}
	ds^2=\left(\eta_{ab}+\frac{\kappa^{n}}{r^n}v_{a}v_{b}\right)d\sigma^a d\sigma^b
	+dr^2+r^2d\Omega_{n+1}^2\,.
\end{equation}
Equivalently, upon introducing the $D$-dimensional null one-form $k_\mu dx^\mu := v_a d\sigma^a$, this can be written in Kerr--Schild form
\begin{equation}
	g_{\mu\nu}=\eta_{\mu\nu}+H(r)k_\mu k_\nu\,, \qquad \text{with}  \qquad H(r)=\kappa^n/r^n\,.
\end{equation}
Since $H$ is harmonic on the transverse $\mathbb{R}^{n+2}$ away from the origin, the geometry solves the vacuum Einstein equations for $r\neq 0$ (for a review, see~\cite{Roche:2022bcz}). In particular, all curvature invariants vanish away from the singularity.

At ideal order only the combination $\kappa^n v_a v_b$ enters the metric. This makes it convenient to package it into the dimensionful null vector $\ell_a := \kappa^{n/2} v_a$, so that the pp-wave term becomes $r^{-n}\ell_a \ell_b$. This is the gravitational counterpart of the ideal-order gauge redundancy discussed in Section~\ref{sec:nullfluids}.

\paragraph{Null energy-momentum tensor from infinity.}
Finally, we may compute the conserved worldvolume energy-momentum tensor associated with~\eqref{eq:dsSchnullideal} directly using the Brown--York prescription~\eqref{eq:BYflat} and the same asymptotic boundary geometry $\mathbb{R}^{1,p}\times {S}^{n+1}$.
Integrating over the transverse sphere and taking $R\to\infty$ gives
\begin{equation}
	\label{eq:nullTabdsSch}
	\mathcal{T}_{ab}^{(0)}=\mathcal{E}v_{a}v_{b}\,, \qquad
	\mathcal{E}=\frac{n\,\Omega_{n+1}}{16\pi G_D}\,\kappa^{n}\,.
\end{equation} 
This is the expected result from taking the ultraboost of the ideal-order energy-momentum tensor~\eqref{eq:TabdsSch}, since the pressure scales as $P\propto r_0^n\to 0$, while $(\varepsilon + P)u_a u_b \sim n\Omega^{n+1}(\gamma^2 r_0^n U_a U_b)/(16\pi G_D)$ remains finite by~\eqref{eq:r0LlimitdsSch} and produces the energy-momentum tensor of null dust.

\subsection{First-order metric}
\label{sec:firstflat}

We now proceed with the construction of the first-derivative corrections to the ideal asymptotically flat pp-wave metric~\eqref{eq:dsSchnullideal} using the blackfold framework. The ideal solution is parametrised by a scale $\kappa$ and a null worldvolume direction $v_a$. In the hydrodynamic regime we promote these collective variables to slowly varying fields on the worldvolume 
\begin{equation}
    \{\kappa,v_a\} \longrightarrow \{\kappa(\sigma),v_a(\sigma)\}\,,
\end{equation}
and organise the expansion by the formal derivative-counting parameter $\epsilon$ as in Section~\ref{sec:AdSbottomup}, with
\begin{equation}
\label{eq:epsilon-flat}
	\partial_a \kappa \sim \mathcal{O}(\epsilon)\,,\qquad \partial_a v_b \sim \mathcal{O}(\epsilon)\,,
	\qquad\text{and}\qquad
	\epsilon\sim \frac{\kappa}{\mathcal{R}}\ll 1\,,
\end{equation}
where $\mathcal{R}$ is the macroscopic variation scale of the collective fields. In a neighbourhood of a point $\sigma=0$ one may Taylor expand as in~\eqref{eq:vkcollvar}. As in the AdS analysis, it is convenient to work locally in coordinates $\sigma^a=(t,x^i,z)$ adapted to the equilibrium null direction, so that at $\epsilon=0$ the null vector lies in the $(t,z)$-plane and the indices $i,j=1,\dots,p-1$ label worldvolume directions transverse to it.

\paragraph{Ansatz and decomposition of Einstein's equations.}
Since the ideal null geometry has no horizon, we do not impose horizon regularity conditions. Instead, we work in a radial gauge and impose asymptotic flatness. We take the first-order corrected metric to be
\begin{equation}
	\label{eq:dsSchnullbottomupf}
	ds^{2}=\left(\eta_{ab}+\frac{\kappa(\sigma)^n}{r^n} v_a(\sigma) v_b(\sigma)\right)d\sigma^ad\sigma^b
	+dr^2+r^2d\Omega_{n+1}^2
	+ f_{\mu\nu}(r) dx^{\mu}dx^{\nu}\ ,
\end{equation}
where $f_{\mu\nu}\sim\mathcal{O}(\epsilon)$ encodes the corrections required for the metric to solve the vacuum Einstein equations
\begin{equation}
\label{eq:vacuum-Einstein-eq}
    R_{\mu\nu} = 0\,,
\end{equation}
once $\kappa$ and $v_a$ vary. At first derivative order it is consistent to take $f_{\mu\nu}$ to depend on $r$ only, since any $\sigma$-dependence enters through first-derivative data multiplying the radial integration functions obtained by solving the ODEs. We focus on intrinsic worldvolume hydrodynamic perturbations and therefore set
\begin{equation}
	f_{\mu\Omega}=0\,,
\end{equation}
so that we do not excite the sector describing extrinsic (or bending) perturbations of the brane (cf.~\cite{Camps:2012hw}). We will work in radial gauge, setting $f_{ra}=0$ at the end. As shown in Appendix~\ref{app:fraremoval}, this is always possible.

Einstein's equations split into radial constraints and dynamical radial ODEs. The constraints are provided by the mixed radial and worldvolume components $R_{ra}=0$, which at first derivative order give
\begin{equation}
	\label{eq:constrainteqns}
	\partial_a(\kappa^n v^a)=\tau_a\dot v^a\,,\qquad h^{ab}\dot v_a=0\,.
\end{equation}
As in the AdS case, we will verify below that~\eqref{eq:constrainteqns} reproduces the ideal-order null fluid equations~\eqref{eq:eomESBP}. The remaining components $R_{\mu\nu}=0$ determine $f_{\hat a\hat b}$ with $\hat a=(a,r)$. Once the constraints~\eqref{eq:constrainteqns} are imposed, the dynamical equations for $f_{\hat a\hat b}$ simplify to homogeneous radial ODEs. Hence the choice $f_{\hat a\hat b}=0$ is a consistent first-order solution, but it does not capture the most general family compatible with asymptotic flatness and our gauge choice. In order to obtain that family we solve the homogeneous ODEs below and keep the allowed homogeneous modes. In particular, imposing asymptotic flatness with the boundary metric fixed to be $\eta_{ab}$ amounts to the requirement that $f_{ab} \sim \mathcal{O}(r^{-n})$ as $r\to\infty$, which sets all terms in $f_{ab}$ at order $r^0$ equal to zero, similar to the nonrenormalisable $r^2$-terms in the AdS analysis of Section~\ref{sec:AdSbottomup}.

\paragraph{Solving the dynamical equations.}
We now solve the $\mu\nu$-components of the vacuum Einstein equations~\eqref{eq:vacuum-Einstein-eq} for the first-order corrections $f_{\hat a \hat b}$. The explicit ODEs are collected in Appendix~\ref{app:dynamicalflat}. For $p\ge 3$, the spatial components $R_{ij}=0$ yield
\begin{equation}
	\label{eq:solfij}
	f_{ij}=-\frac{c^{(1)}_{ij}}{n r^{n}}
    \,,
\end{equation}
where $c^{(1,2)}_{ij}(\sigma)\sim\mathcal{O}(\epsilon)$ are radial integration functions. This solution is absent when $p=1$, and when $p=2$ there are no off-diagonal components.

The components $R_{ti}=0$ and $R_{zi}=0$ are coupled and give
\begin{equation}
    \label{eq:solftifzi}
	\begin{split}
		f_{ti}&=
        -\frac{c^{(2)}_{ti}}{n r^n}
		+\frac{c^{(1)}_{iz}+c^{(2)}_{iz} r^n}{n r^n \kappa^{n}}
		-\frac{c^{(1)}_{iz}}{2n r^{2n}}\,,\\
		f_{iz}&=
        -\frac{c^{(2)}_{iz}}{n r^n}-\frac{c^{(1)}_{iz}}{2n r^{2n}}\,,
	\end{split}
\end{equation}
with the same constraint~\eqref{eq:cond1} on the integration functions we found for AdS. Solving the coupled system consisting of the equations $R_{tt}=R_{tz}=R_{zz}=R_{rr}=0$ yields
\begin{equation}
	\label{eq:solftzr}
	\begin{split}
		f_{tt}&=c^{(1)}_{tt}
		+\frac{\tilde c^{(4)}}{n\kappa^n}
		+\frac{\tilde c^{(3)}-\kappa^n (c^{(1)}_{zz}+2 \tilde c^{(4)})}{n r^n \kappa^n}-\frac{\tilde c^{(3)}-\kappa^n \delta^{ij}c^{(1)}_{ij}}{2n r^{2n}}	+\frac{\kappa^n \tilde c^{(3)}}{nc_n r^{3n}}\,,\\
		f_{tz}&=-\frac{c^{(1)}_{zz}+ \tilde c^{(4)}}{n r^n}
		+\frac{\kappa^n \delta^{ij}c^{(1)}_{ij}}{2n r^{2n}}
		+\frac{\kappa^n \tilde c^{(3)}}{nc_n r^{3n}}\,,\\
		f_{zz}&=-\frac{c^{(1)}_{zz}}{n r^n}
		+\frac{\tilde c^{(3)}+\kappa^n \delta^{ij}c^{(1)}_{ij}}{2n r^{2n}}
		+\frac{\kappa^n \tilde c^{(3)}}{nc_n r^{3n}}\,,\\
		f_{rr}&=c_{rr}-\frac{2c^{(1)}_{zz}+\tilde c^{(4)}}{n\kappa^{n}}
		+\frac{\kappa^n \big(n c^{(1)}_{tt} \kappa^n+ \delta^{ij} c^{(1)}_{ij}-\tilde c^{(4)}\big)+\tilde c^{(3)}}{n r^n \kappa^n}
		-\frac{\tilde c^{(3)}}{2n r^{2n}}\,,
	\end{split}
\end{equation} with the constraint $c^{(1)}_{tt}+\tilde c^{(4)}/(n\kappa^n)=0$ and \begin{equation}
    c_{n}=\frac{12(n+1)}{n+2}\ ,
\end{equation} is a dimension-dependent parameter\footnote{As seen for the AdS case, this parameter has been checked for specific cases. In particular, for $(p,n)\in\{1,2,3,4\}$.}. Moreover, the angular equation $R_{\Omega\Omega}=0$ then fixes
\begin{equation}
	c_{rr}=\frac{2c^{(1)}_{zz}+\tilde c^{(4)}}{n\kappa^{n}}\,.
\end{equation}
Finally, just as for AdS, we impose radial gauge by setting
\begin{equation}
	f_{ra}=0\,,
\end{equation}
which is discussed in detail in Appendix~\ref{app:fraremoval}. 

\paragraph{Energy-momentum tensor and matching with null fluids.}
We now determine which choices of integration functions correspond to hydrodynamic perturbations of a null fluid. We compute the quasilocal Brown--York stress tensor on a timelike hypersurface at constant $r$, integrate over the transverse ${S}^{n+1}$, and take the large-radius limit. In blackfold language, this corresponds to placing the ``screen'' in the overlap region
\begin{equation}
	\kappa \ll r \ll \mathcal{R}\,,
\end{equation}
where the derivative expansion is valid and the integration over the transverse sphere produces an effective worldvolume stress tensor. 

Finiteness of the stress tensor combined with the absence of angular components enforce $\tilde c^{(3)}=0$. Defining the first-order correction by subtracting the ideal contribution, we obtain after imposing~\eqref{eq:cond1}
\begin{equation}
	\label{eq:cT1}
	\begin{split}
		\tilde{\mathcal{T}}_{tt} &= \frac{1}{n}\Big(n c^{(1)}_{tt} \kappa^n+c^{(1)}_{ii}-n c^{(1)}_{zz}-(2n+1) \tilde c^{(4)}\Big)\,,\\
		\tilde{\mathcal{T}}_{tz}&= -\big(c^{(1)}_{zz}+\tilde c^{(4)}\big)\,,\\
		\tilde{\mathcal{T}}_{ti} &= -\Big(c^{(2)}_{iz}-\frac{ c^{(1)}_{iz}}{\kappa^n}\Big)\,,\\
		\tilde{\mathcal{T}}_{zi} &= -c^{(2)}_{iz}\,,\\
		\tilde{\mathcal{T}}_{ij}&=-c^{(1)}_{ij}-\frac{\delta_{ij}}{n}\Big(n c^{(1)}_{tt} \kappa^n+ \delta^{kl} c^{(1)}_{kl}- \tilde c^{(4)}\Big)\,.
	\end{split}
\end{equation}
To match the bulk solution to null fluid data, we now express the remaining $\sigma$-dependent
radial integration functions in terms of first-derivative structures built from the collective fields
$(\kappa(\sigma),v_a(\sigma))$, in direct analogy with the AlAdS analysis of Section~\ref{sec:AdSbottomup}. As in that discussion, it is convenient to work locally in an adapted patch where, at ideal order, the null direction lies in the $(t,z)$-plane. In such coordinates, we may use $\partial_a\kappa$ and $\partial_a v_b$ as a component basis, while keeping in mind that covariance must be restored in the final expressions: any non-covariant bookkeeping terms introduced at intermediate steps must either vanish or recombine into covariant null-fluid data such as $v^a\partial_a\kappa$, $\vartheta=\partial_a v^a$, $\dot v_a:=v^b\partial_b v_a$, $\partial_{(a}v_{b)}$, and the shear $\varsigma_{ab}$.

We therefore parametrise the scalar, vector, and tensor integration functions by writing them as linear
combinations of the available first-derivative structures, with coefficient functions $a_\bullet(\kappa)$.
These $a_\bullet$ are dimensionful functions of $\kappa$, introduced purely to keep track of the most general first-order dependence compatible with the derivative counting. Concretely, in the scalar sector we take
\begin{equation}
	\label{eq:three-cs}
	\begin{split}
		c^{(1)}_{tt}&=a_{1}\vartheta+a_2\frac{v^a\partial_a\kappa}{\kappa}
		+\frac{2(a_3\partial_t+a_4\partial_z)\kappa}{\kappa}\, , \\
		c^{(1)}_{zz}&=a_{5}\vartheta+a_6\frac{v^a\partial_a\kappa}{\kappa}
		+\frac{2(a_7\partial_t+a_8\partial_z)\kappa}{\kappa}\, ,\\
		\tilde c^{(4)}&=a_{9}\vartheta+a_{10}\frac{v^a\partial_a\kappa}{\kappa}
		+\frac{2(a_{11}\partial_t+a_{12}\partial_z)\kappa}{\kappa}\, ,
	\end{split}
\end{equation}
where the terms involving $\partial_t\kappa$ and $\partial_z\kappa$ are non-covariant artefacts of working
in adapted coordinates. Requiring that the final metric correction $f_{\hat a\hat b}$ and the quasi-local energy-momentum tensor $\tilde{\mathcal T}_{ab}$ can be written covariantly forces these artefacts to either disappear or to recombine into $v^a\partial_a\kappa$, exactly as in the AlAdS case.

In the vector sector the coupled radial solutions impose the condition~\eqref{eq:cond1}, which restricts the
allowed combinations of integration functions. A convenient parametrisation that automatically respects this constraint is
\begin{equation}
	\begin{split}
		c^{(2)}_{ti}&=-2\Big(a_{13}v_{(t}\dot v_{i)}+a_{14}\frac{v_{(t}\partial_{i)}\kappa}{\kappa}
		+a_{15}\partial_{(z}v_{i)}\Big)\, , \\
		c^{(2)}_{zi}&=-2\Big(a_{13}v_{(z}\dot v_{i)}+a_{14}\frac{v_{(z}\partial_{i)}\kappa}{\kappa}
		+a_{15}\partial_{(z}v_{i)}\Big)\, , \\ 
		c^{(1)}_{iz}&=\kappa^{n}a_{15}\dot v_i\, ,
	\end{split}
\end{equation}
where $\partial_z v_i$ is again kept only as a bookkeeping device; in the final covariant answer it is absorbed
into $\partial_{(a}v_{b)}$ and, after subtracting the trace, into $\varsigma_{ab}$. Finally, in the symmetric
tensor sector we take
\begin{equation}
	c_{ij}^{(1)}=-2\left(a_{16}\partial_{(i}v_{j)}+a_{17}\frac{v_{(i}\partial_{j)}\kappa}{\kappa}
	+a_{18}v_{(i}v_{j)}\vartheta\right)\, .
\end{equation}
Altogether this introduces $18$ independent coefficient functions $a_\bullet(\kappa)$. 
They again are fixed by requiring that (i) the metric correction $f_{\hat a\hat b}$ is compatible with our gauge choice, 
(ii) the quasi-local energy-momentum tensor on the screen $\kappa\ll r\ll \mathcal R$ is finite and has the expected tensor
structure along the worldvolume, and (iii) both $f_{\hat a\hat b}$ and $\tilde{\mathcal T}_{ab}$ can be expressed covariantly and symmetrically in terms of first-order null fluid data. Imposing these conditions yields the relations
\begin{equation}\label{eq:afixflat}
	\begin{split}
		a_{1,3,4,7,9,11,12}&=0\, , \qquad
		a_{15}=a_{16}=\kappa^{n+1}\tilde a_\eta=\kappa^{n+1}\frac{n\tilde a_1}{2}\, , \qquad
		a_5=2a_{18}=2\kappa^{n+1}\tilde a_2\, ,\\
		a_{13}&=\kappa^{n+1}\tilde a_3\, , \qquad
		a_6=2\kappa^{n+1}\tilde a_4\, , \qquad
		a_8=a_{14}=a_{17}=\kappa^{n+1}\tilde a_5\, ,\\
		a_{10}&=-\frac{n\kappa^n a_2}{2}=-\kappa^{n+1}\tilde a_6\, , \qquad
		\tilde a_6=\frac{2}{n}\tilde a_5\, ,
	\end{split}
\end{equation}
where it is convenient to package the remaining freedom into the dimensionless constants $\tilde a_\bullet$,
defined exactly as in Section~\ref{sec:AdSbottomup} but with $n$ replacing $d$.

\paragraph{Resulting first-order geometry and energy-momentum tensor.}
Combining our above findings,\footnote{The results presented here have been explicitly checked for $(p,n)\in\{1,2,3,4\}$.} the first-order corrected asymptotically flat metric is~\eqref{eq:dsSchnullbottomupf} with
\begin{equation}
	\label{eq:mostgenfmunu}
	\begin{split}
		f_{ab}&=-\frac{2\kappa^{n+1}}{n r^n}\Big[\Big(\tilde a_2\vartheta +\tilde a_4 \frac{v^c\partial_c\kappa}{\kappa}\Big)v_av_b
		+\tilde a_3v_{(a}\dot v_{b)}
		+\tilde a_6\frac{n\,v_{(a}\partial_{b)}\kappa}{2\kappa}
		+\tilde a_\eta\,\partial_{(a}v_{b)}\Big]\\
        &\quad \,+\frac{\kappa^{2n+1}}{n r^{2n}}\tilde a_\eta\big(\vartheta v_a v_b-v_{(a}\dot v_{b)} \big)\,,
		\\
		f_{rr}&=\frac{2\kappa^{n+1}}{n r^n}\Big(\tilde a_\eta\vartheta +\tilde a_6 \frac{n\,v^a\partial_a\kappa}{2\kappa}\Big)\,.
	\end{split}
\end{equation}
We have checked explicitly that the Kretschmann scalar remains $K_r=0+\mathcal{O}(\epsilon^2)$, so that the spacetime remains locally flat to this order and that no horizon is generated at first derivative order.

The effective worldvolume energy-momentum tensor takes the null fluid form
\begin{equation}
	\label{eq:mostgenstress}
	\mathcal{T}^{ab}_{(1)}=\left(\rho_1\vartheta+\rho_6\frac{v^c\partial_c\kappa}{\kappa}\right) \eta^{ab}
	+2v^{(a}\left(\rho_2\vartheta v^{b)}+\rho_3\dot{v}^{b)}+\rho_4v^{b)}\frac{v^c\partial_c\kappa}{\kappa}
	+\rho_5\frac{\partial^{b)}\kappa}{\kappa}\right)-2\eta\varsigma^{ab}\,,
\end{equation}
with
\begin{equation}
	\rho_\bullet = -\frac{\Omega_{n+1}}{16\pi G}\kappa^{n+1}\tilde a_\bullet \,,\qquad
	\eta=\frac{\Omega_{n+1}}{16\pi G}\kappa^{n+1}\tilde a_\eta\,,\qquad
	\rho_1=-\frac{2}{n}\eta\,,\qquad \rho_6=\frac{2}{n}\rho_5\,.
\end{equation}
Thus $\mathcal{T}^{ab}_{(1)}$ matches~\eqref{eq:firstnullTgen} subject to constraints among transport coefficients imposed by the gravity equations, just as in the timelike blackfold setting, and conservation of $\mathcal{T}^{ab}$ reproduces the constraint equations~\eqref{eq:constrainteqns}, confirming that the bulk constraints are precisely the null hydrodynamic equations.

\paragraph{Bulk frame transformations.}
In both the hydrodynamic and gravity descriptions, the first-order null fluid constitutive data admit frame redefinitions, which shift $(\kappa,v_a)$ by terms that are $\mathcal{O}(\epsilon)$ and correspondingly reorganise the first-derivative structures appearing in both the energy-momentum tensor and the metric. In particular, in the constant-pressure subsector, where $\rho_2,\rho_3,\rho_4$
are redundant, one may choose a convenient representative in which these terms are absent.
Following the general construction in Appendix~\ref{app:frames}, the choice
\begin{equation}
	\label{eq:frametflat}
	\delta g_{ab}(\delta v,\delta\kappa)
	=\delta g_{ab}\left(\tilde a_3\,\frac{\kappa}{n}\dot v\,,\,
	\frac{2\kappa^2}{n^2}\Big(\tilde a_2\,\vartheta+\tilde a_4\,\frac{v^c\partial_c\kappa}{\kappa}\Big)\right)\,,
\end{equation}
implements the corresponding frame redefinition. Using~\eqref{eq:pertmetric} shows that this transformation eliminates the pieces of $f_{\hat a\hat b}$ proportional to $\tilde a_2$, $\tilde a_3$, and $\tilde a_4$, and correspondingly removes the $(\rho_2,\rho_3,\rho_4)$ structures from the boundary energy-momentum tensor, producing the form of the first-order energy-momentum tensor in~\eqref{eq:firstnullgenP}.

Finally, in the special case $\rho_1=\rho_6=\eta=0$, i.e., $\tilde a_1=\tilde a_\eta=\tilde a_6=0$, the same transformation is sufficient to bring the first-order metric into a gauge where
\begin{equation}
	f_{\mu\nu}=0\,,
\end{equation}
so that the geometry reduces to~\eqref{eq:dsSchnullbottomupf} at $\mathcal{O}(\epsilon)$.

\subsection{Specific frames from null limit}
\label{sec:nullflat}

In \cite{Camps:2010br}, the first-order hydrodynamic perturbations of the boosted Schwarzschild black $p$-brane~\eqref{eq:dsSch} were constructed by promoting the collective variables $(r_0,u_a)$ to slowly varying worldvolume fields and solving the vacuum Einstein equations to first order in derivatives. For the reader’s convenience, we summarise the resulting first-order metric and the associated effective energy-momentum tensor in Appendix~\ref{app:firstflat}.

The blackfold derivative expansion requires a separation of scales between the variation scale $\mathcal{R}$ of the collective fields and the brane thickness $r_0$,
\begin{equation}
\label{eq:sep-of-scales}
	r_0 \ll \mathcal{R} \qquad \iff \qquad 	\mathcal{R}T \gg 1\,,
\end{equation}
with $T=\frac{n}{4\pi r_0}$ (cf.~\eqref{eq:Tr0dsSch}). Since the null limit studied in Section~\ref{sec:idealflat} corresponds to $r_0\to 0$, and hence $T\to\infty$, the separation of scales required for the blackfold description~\eqref{eq:sep-of-scales} is automatically satisfied, since $\mathcal{R}T\to \infty$. In order to obtain a finite and non-trivial null limit of the first-derivative corrections, one must in addition take a double-scaling limit in which the first-derivative data are scaled with an appropriate power of the Lorentz factor, exactly as in Section~\ref{sec:nulllimitofrelfluids}, and~\eqref{eq:r0LlimitdsSch} tells us that
\begin{equation}
    \label{eq:gamma-bf-scaling}
    \gamma \sim \mathcal{O}\big(r_0^{-n/2}\big)\,.
\end{equation}
We now show that, in two different timelike frames, the null limit of the timelike first-order blackfold metrics reproduces certain special cases of the bottom-up null solution~\eqref{eq:dsSchnullbottomupf} obtained in Section~\ref{sec:asyflatnullsol}.

\subsubsection*{Landau frame}

Taking the null limit directly in the timelike Landau frame produces a special case of the general first-order null solution~\eqref{eq:mostgenstress}, namely the one for which
\begin{equation}
	\rho_{1,4,5,6}=0\,,\qquad \eta=0\,.
\end{equation}
Equivalently, on the gravity side, this corresponds to the restriction $\tilde a_{4,5,6,\eta}=0$ in~\eqref{eq:mostgenfmunu}, so that
\begin{equation}
	\label{eq:fabtimelandau}
	f_{ab}=-\frac{2\kappa^{n+1}}{n r^n}\left(\tilde a_2\vartheta v_a v_b+\tilde a_3 v_{(a}\dot v_{b)}\right)\,, \qquad f_{rr}=0\,.
\end{equation}
We now obtain~\eqref{eq:fabtimelandau} from the null limit of the timelike first-order blackfold solution and fix $(\tilde a_2,\tilde a_3)$.

\paragraph{Hydrodynamic scaling.}
Starting from the timelike first-order stress tensor in Landau frame~\eqref{eq:Tab1nonnull} written in the form~\eqref{eq:T1nonull}, and using the scaling limit~\eqref{eq:limitT1landau}, one finds
\begin{equation}
	\label{eq:speclimLandau}
	\begin{split}
		\frac{2\hat\eta}{n+1}\gamma^{\alpha+3}	&\sim \frac{\Omega_{n+1}}{8(n+1)\pi G_D}r_0^{n+1-\frac{n}{2}(3+\alpha)} \rightsquigarrow \frac{\Omega_{n+1}}{8(n+1)\pi G_D}\kappa^{n+1} =: 2\rho_2\,,	\\
		2\hat\eta\gamma^{\alpha+3}	&\sim \frac{\Omega_{n+1}}{8\pi G_D}r_0^{n+1-\frac{n}{2}(3+\alpha)}  \rightsquigarrow	\frac{\Omega_{n+1}}{8\pi G_D}\kappa^{n+1}=: 2\rho_3\,,
	\end{split}
\end{equation}
which is finite and non-trivial if
\begin{equation}
	\alpha=\frac{2-n}{n}\,.
\end{equation}
In particular, the limiting coefficients satisfy $\rho_3=(n+1)\rho_2=\frac{\Omega_{n+1}}{16\pi G_D}\kappa^{n+1}$.

\paragraph{Gravitational limit.}
On the bulk side we apply the same null limit directly to the timelike first-order blackfold metric (in the Schwarzschild-like gauge used in Appendix~\ref{app:firstflat}). For illustration, consider the contribution of the $f_1$ term in~\eqref{eq:fmunuflat}:
\begin{equation}
	\label{eq:limitafixing}
	\theta u_a u_b f_1\sim	U_aU_b\,\nabla_\alpha U^\alpha r_0^{-\frac{n}{2}(3+\alpha)}f_1 \rightsquigarrow	-\frac{2\kappa^{n+1}}{n(n+1)r^n}v_a v_b\vartheta\,,
\end{equation}
where we used~\eqref{eq:gamma-bf-scaling} and the same value $\alpha=(2-n)/n$ found from the hydrodynamic scaling. Repeating this for the remaining first-order structures produces the null metric~\eqref{eq:dsSchnullbottomupf} with
\begin{equation}
	\label{eq:fmununullflat}
	f_{ab}=-\frac{2\kappa^{n+1}}{n r^n}\left(\frac{1}{n+1} \vartheta v_a v_b+ v_{(a}\dot v_{b)}\right)\,, \qquad f_{rr}=0\,,
\end{equation}
and all other components vanishing after imposing the gauge $f_{ra} = 0$ (see Appendix~\ref{app:fraremoval} for details). Comparing with~\eqref{eq:fabtimelandau} fixes
\begin{equation}
	\tilde a_2=\frac{1}{n+1}\,, \qquad 	\tilde a_3=1\,.
\end{equation}
This matches the hydrodynamic limit~\eqref{eq:speclimLandau} and reproduces the expected null fluid energy-momentum tensor in Landau frame.

Finally, the frame freedom discussed in Section~\ref{sec:nullfluids} allows us to remove the surviving $(\rho_2,\rho_3)$ structures. On the gravity side, the corresponding reparametrisation of the collective fields may be chosen so that the induced first-order change of the metric cancels the $(\tilde a_2,\tilde a_3)$ terms, yielding a representative with $f_{\mu\nu}=0$ and hence $\mathcal{T}^{ab}_{(1)}=0$.

\subsubsection*{Non-thermodynamic frame}

This case again corresponds to a special of the general first-order solution found above.
At the level of the null fluid energy-momentum tensor, this is obtained by setting $\rho_{2,3,4}=0$ in~\eqref{eq:mostgenstress}. Equivalently, using $\tilde a_1=2\tilde a_\eta/n$ and setting $\tilde a_{2,3,4}=0$ in~\eqref{eq:mostgenfmunu}, the metric perturbation takes the form
\begin{equation}
	\label{eq:fmunuspecnullLandau}
	f_{ab}=\tilde a_\eta\,\kappa^{n+1}\left(
	-\frac{2}{n r^n}\varsigma_{ab}
	+\frac{\kappa^n}{n r^{2n}}\left(\vartheta v_a v_b-v_{(a}\dot v_{b)}\right)\right)\,,
	\qquad
	f_{rr}=\frac{2\tilde a_\eta\,\kappa^{n+1}}{n r^n}\,\vartheta\,.
\end{equation}
We now show that the same structure follows from the null limit of the first-order perturbed black brane after a
timelike frame redefinition. We next take the null limit after first performing a timelike frame transformation away from the Landau frame, so that the timelike energy-momentum tensor takes the non-thermodynamic form~\eqref{eq:nullLandauT1}. At the level of the timelike collective variables, a convenient choice is
\begin{equation}
	\label{eq:frametransfflat}
	\delta u_a=\frac{r_0}{n} \dot u_a\,,\qquad 	\delta r_0=\frac{2}{n^2(n+1)} r_0^2\,\vartheta\,,
\end{equation}
implying that~\eqref{eq:Tab1nonnull} takes the form
\begin{equation}
	T_{ab}=T_{ab}^{(0)}-2\hat\eta\left(\partial_{(a}u_{b)}+\frac{\vartheta}{n} \eta_{ab}\right)\,.
\end{equation}

\paragraph{Hydrodynamic scaling.}
Taking the null limit of the first-order terms in this frame (cf.~\eqref{eq:limitT1nulllandu}) gives
\begin{equation}
	\label{eq:limnullLandau}
	\begin{split}
		2\hat\eta \gamma^{\alpha+1} &\sim \frac{\Omega_{n+1}}{8\pi G_D} r_0^{n+1-\frac{n}{2}(1+\alpha)}  \rightsquigarrow 	\frac{\Omega_{n+1}}{16\pi G_D} \kappa^{n+1} =: 2\eta\,, \\
		\frac{2}{n}\hat\eta \gamma^{\alpha+1} &\sim \frac{\Omega_{n+1}}{8n\pi G_D} r_0^{n+1-\frac{n}{2}(1+\alpha)} \rightsquigarrow \frac{\Omega_{n+1}}{8n\pi G_D} \kappa^{n+1} =: 2\rho_1\,,
	\end{split}
\end{equation}
which is finite and non-trivial provided 
\begin{equation}
    \alpha = \frac{n+2}{n}\,.
\end{equation}

\paragraph{Gravitational limit.}
On the bulk side we first implement the corresponding frame transformation of the timelike first-order metric (cf.~Appendix~\ref{app:frames}) and then fix radial gauge by eliminating the mixed components $f_{ra}$ as discussed in Appendix~\ref{app:fraremoval}. The resulting first-order coefficients can be written as shifted functions $\tilde f_{\mu\nu}$
\begin{equation}\begin{split}
	\label{eq:tildefmunu}
	\tilde f_{ab}&=f_{ab}+\frac{r_0^{n+1}}{r^n}\left(\frac{2\theta}{n(n+1)}u_a u_b+\frac{2}{n}u_{(a}\dot u_{b)}\right)\,,\\ 
    \tilde f_{rr}&=f_{rr}+\frac{r^2 r_0^{n+1}\theta}{(n+1)(r^n-r_0^n)^2}\,,\qquad 	\tilde f_{ra}=0\,.
\end{split}\end{equation}
Taking the null limit with scaling exponent $\alpha=(n+2)/n$ and using again~\eqref{eq:gamma-bf-scaling}, we find a non-trivial $\mathcal{O}(r^{-2n})$ contribution:
\begin{equation}
	\label{eq:nulllandauf1}
	\theta u_au_b\,\tilde f_1 \sim 
	U_aU_b\,\nabla_\alpha U^\alpha 	r_0^{-\frac{n}{2}(1+\alpha)} \tilde f_1  \rightsquigarrow 	\frac{\kappa^{2n+1}}{2r^{2n}} v_a v_b \vartheta\,.
\end{equation}
The remaining terms yield, mutatis mutandis, the null metric~\eqref{eq:dsSchnullbottomupf} with
\begin{equation}
	\label{eq:fabnonthermoflat}
	f_{ab}=\kappa^{n+1}\left(
	-\frac{2}{n r^n} \varsigma_{ab} +\frac{\kappa^n}{n r^{2n}}\left(\vartheta v_a v_b-v_{(a}\dot v_{b)}\right)\right)\,, \qquad 	f_{rr}=\frac{2\kappa^{n+1}}{n r^n}\vartheta\,,
\end{equation}
which matches the restricted bottom-up class~\eqref{eq:fmunuspecnullLandau} upon identifying $\tilde a_\eta=1$. In other words, the limiting null energy-momentum tensor has $\rho_{2,3,4}=0$ and satisfies $\eta=-\frac{n}{2}\rho_1=-\frac{\Omega_{n+1}}{16\pi G_D} \kappa^{n+1}$, in agreement with~\eqref{eq:limnullLandau}.

\section{Discussion}\label{sec:discussion}
In this work we performed long-wavelength perturbations of Kaigorodov spacetimes in AdS and obtained new inhomogeneous pp-wave spacetimes by allowing the pp-wave amplitude and pp-wave direction to be slowly varying functions of boundary coordinates. From the boundary point of view, the pp-wave amplitude corresponds to momentum density and pp-wave direction to null fluid velocity. We showed in particular that the long-wavelength perturbations of the Kaigorodov metrics translate into a null hydrodynamic expansion as spelled out in \cite{Armas:2025uyv} from the viewpoint of the CFT. Particular classes of these inhomogeneous metrics can be obtained from an ultra-relativistic limit of boosted planar AdS black holes in which the boost velocity tends to infinity and the temperature vanishes. Therefore they can be interpreted as null states in the CFT, obtainable from a zero-temperature and large-momentum limit of finite temperature states. We also derived the corresponding results for perturbations of asymptotically flat pp-wave solutions, some of which can be obtained from infinitely boosted black $p$-branes (blackfolds). Let us emphasise that the derivative expansion used here should not be understood as ordinary finite-temperature hydrodynamics of the limiting state. Rather, it is an effective long-wavelength expansion of energy-momentum tensor configurations that are controlled by the finite scale $\kappa$. When such configurations arise from a double-scaling limit of thermal states, this scale is inherited from the parent fluid. In the more general intrinsic description, the scale $\kappa$ instead sets the local scale associated with the chosen normalisation of the null fluid velocity.

The inhomogenoeus pp-wave metrics we found by considering first-order gradient corrections of pp-wave amplitude and direction are parametrised by 7 parameters $\tilde a_\bullet$, of which only 5 are independent. It is possible to perform field redefinitions (or equivalently null frame transformations from the boundary point of view) to remove three of these parameters leaving only two independent parameters. The transport coefficients associated with these perturbed metrics are determined in terms of the scale $\kappa$ up to numerical factors, and the ratio of the remaining transport coefficients is completely fixed and given by
\begin{equation}\label{eq:ratiofirst}
    \frac{\rho_1}{\eta}=\frac{2}{d} \ \ \ \&\ \ \  \frac{\rho_6}{\rho_5}=-\frac{2}{d}\, .
\end{equation}
These ratios are compatible with the conformal nature of the respective energy-momentum tensors at the boundary. In the case of asymptotically flat pp-waves, the ratios pick up a minus sign and $d$ is replaced by $n$. We thus see that gravity picks out a particular realisation of the general theory of null fluids discussed in Section~\ref{sec:nullhydrodynamics}. Let us also clarify the status of the first-order coefficients. The gravitational calculation fixes the allowed tensor structures and the radial dependence, and it imposes constraints among the coefficients via Einstein's equations. Any coefficient that can be removed by residual gauge transformations or by redefinitions of the null fluid variables is not physical. The frame invariant combinations that remain make up the transport data of the null fluid. However, because the pp-wave geometries considered here do not possess a regular finite-temperature horizon, there is no horizon regularity condition that fixes all of their values. The remaining homogeneous normalisable modes should therefore be regarded as state-dependent data, rather than as universal transport coefficients predicted by gravity alone. A null limit of a regular timelike black brane remembers horizon regularity which supplies an additional prescription and hence selects a particular subset of these coefficients.

It would be interesting to explore various extensions of the ideas presented here. For instance, there are more general classes of AdS-pp-wave metrics that include Kaigorodov metrics as a special case \cite{Brecher:2000pa} which could constitute the starting point for long wavelength perturbations. In this direction it could be relevant to check whether the ratios \eqref{eq:ratiofirst} are universal for this larger class of metrics. Another interesting direction is to extend null hydrodynamics to include U(1) charges and U(1) anomalies and to look at their gravitational duals by considering pp-wave geometries in both ungauged and gauged minimal supergravities \cite{Gauntlett:2003fk, Kerimo:2004gz}, or by taking ultra-relativistic limits of R-charged planar AdS black holes \cite{Erdmenger:2008rm,Banerjee:2008th}. A similar line of thought can be pursued for black $p$-branes carrying higher-form charges which could lead to novel and interesting regimes of magnetohydrodynamics. We intend to pursue some of these ideas in the future.

\section*{Acknowledgments}
\label{sec:acknowledgments}
We are grateful to Matthias Blau, Jan de Boer, Pawel Caputa, José Figueroa-O'Farrill, Jelle Hartong, and Niels Obers for useful discussions. JA is partly funded by the Dutch
Institute for Emergent Phenomena (DIEP) cluster at the
University of Amsterdam via the DIEP programme Foundations and Applications of Emergence (FAEME) and
the national NWA consortium Emergence At All Scales
(EAAS). EH is grateful to Harvard University for hospitality during the final stages of this project. The work of EH is supported by Carlsberg Foundation grant CF24-1656. The Center of Gravity is a Center of Excellence funded by the Danish National Research Foundation under grant No.~184. GPN is partly supported by the Tertiary Education Scholarship Scheme (TESS).

\appendix
\section{First-order black brane metrics}
\label{app:firstordermetrics}
In this appendix we review the blackfold and fluid/gravity metrics that we used throughout this work as a starting point for taking ultra-relativistic limits.

\subsection{Asymptotically flat solutions}
\label{app:firstflat}

This appendix summarises the timelike first-order blackfold solution of~\cite{Camps:2010br} for a neutral boosted Schwarzschild black $p$-brane in
\begin{equation}
	D=p+n+3\,,
\end{equation}
spacetime dimensions, with worldvolume coordinates $\sigma^a$ ($a=0,\dots,p$) and transverse sphere $\mathbb S^{n+1}$.
We work in a derivative expansion controlled by a small parameter (cf.~\eqref{eq:epsilon-flat})
\begin{equation}
	\epsilon \sim \frac{r_0}{\mathcal{R}} \ll 1\,,
\end{equation}
where $r_0(\sigma)$ is the local brane thickness and $\mathcal{R}$ is the characteristic length scale of variations of the collective fields.

At zeroth order in $\epsilon$, the metric is that of the boosted black brane~\eqref{eq:dsSch}. At first order in $\epsilon$, we must take into the account the (slow) variation of $r_0(\sigma)$ and $u^a(\sigma)$, which requires us to add a $\mathcal{O}(\epsilon)$ correction to the metric to preserve the vacuum Einstein equations. In the Schwarzschild-like gauge used in~\cite{Camps:2010br}, the first-order solution can be written as
\begin{equation}
	\label{eq:dsSchnonnullfirst}
	ds^{2}=\left(\eta_{ab}+\frac{r_0^n(\sigma)}{r^n}u_{a}(\sigma)u_{b}(\sigma)\right)d\sigma^a d\sigma^b
	+\frac{dr^2}{f(r)}
	+r^2 d\Omega_{n+1}^2
	+ f_{\mu\nu}(r)\,dx^\mu dx^\nu\,,
\end{equation}
where
\begin{equation}
	f(r)=1-\left(\frac{r_0(\sigma)}{r}\right)^n~,
\end{equation}
is the blackening factor and $f_{\mu\nu}(r)=\mathcal{O}(\epsilon)$ is chosen so that~\eqref{eq:dsSchnonnullfirst} solves the vacuum Einstein equations to first order in $\epsilon$. 

The correction $f_{\mu\nu}$ decomposes into scalar and tensor sectors built from first derivatives of $u_a$ and $r_0$. Recall that
\begin{equation}
	\theta := \partial_a u^a\,, \qquad \dot u_a := u^b \partial_b u_a\,, \qquad
	\Delta_{ab}:=\eta_{ab}+u_a u_b\,,
\end{equation}
with the shear tensor given by
\begin{equation}
	\label{eq:sigmanonnull}
	\sigma_{ab}:=\Delta_a^{\ c}\Delta_b^{\ d}\partial_{(c}u_{d)}-\frac{1}{p}\theta\,\Delta_{ab}\,,
\end{equation}
where indices are raised/lowered with the worldvolume Minkowski metric $\eta_{ab}$. In terms of these first-order structures, the correction $f_{\mu\nu}$ may be written as
\begin{subequations}
	\label{eq:fmunuflat}
	\begin{align}
		f_{ab}(r)	&=\theta u_a u_b\,f_{1}(r)
		+\left(\sigma_{ab}+\frac{1}{p}\theta \Delta_{ab}\right)f_{2}(r)\,,
		\\
		f_{ar}(r)	&=\theta u_{a}\,f_{3}(r)+\dot{u}_{a}\,f_{4}(r)\,,
		\label{eq:fra-flat}
		\\
		f_{rr}(r)	&=\theta f(r)^{-1}f_{5}(r)\,,
	\end{align}
\end{subequations}
with (cf.~Eqs.~(6.11)--(6.16) of~\cite{Camps:2010br}) 
\begin{equation}
	\label{eq:flatfi}
	\begin{split}
		f_{1}(r)
		&=\frac{r_{0}}{n(n+1)} \Big(2-(n+2)\Big(\frac{r_0}{r}\Big)^n\Big) \log f(r)\,,\\
		f_{2}(r) &=\frac{2r_{0}}{n} \log f(r)\,, \\
		f_{3}(r) &=\frac{r_{0}}{(n+1) f(r)}
		\Bigg[
		\Big(\frac{n+1}{n}\Big(\frac{r_0}{r}\Big)^n-\frac{1}{n}\Big)\log f(r)
		-\Big(\frac{r_0}{r}\Big)^n\Big(\frac{n r_{*}(r)}{r_{0}}+1\Big)
		\Bigg] +\delta_{n,1} r_0 \,, \\
		f_{4}(r) &=\frac{r_{*}(r)-r}{f(r)}-\delta_{n,1} r_{0}\log\Big(\frac{r_{0}}{r}\Big)\,, \\
		f_{5}(r) &=\frac{r^{-n}r_{0}^{n+1}}{(n+1) f(r)}
		\Big(2-\log f(r)\Big)\,,
	\end{split}
\end{equation}
where $r_{*}(r)$ is the tortoise coordinate
\begin{equation}
	\label{eq:rstar-flat}
	r_{*}(r)=\int^{r}\frac{dr'}{f(r')}\,.
\end{equation}
The terms involving $\delta_{n,1}$ are there to capture the logarithmic behaviour that arises in the integral~\eqref{eq:rstar-flat} when $n=1$~\cite{Camps:2010br}. The null limit of the metric~\eqref{eq:dsSchnonnullfirst} was studied in Section~\ref{sec:nullflat}. 

Next, we consider the energy-momentum tensor defined on a timelike constant-$r$ hypersurface in the weak-field regime
\begin{equation}
	r_0 \ll r \ll \mathcal{R}\,,
\end{equation}
where we may integrate over the transverse sphere $\mathbb S^{n+1}$ to obtain an effective $(p+1)$-dimensional energy-momentum tensor on the worldvolume, which to first order takes the form
\begin{equation}
	\label{eq:Tab1nonnull}
	T_{ab}=T_{ab}^{(0)}-2\hat\eta \sigma_{ab}-\zeta \theta\Delta_{ab}\,,
	\qquad
	\zeta=2\left(\frac{1}{p} + \frac{1}{n+1}\right)\hat\eta\,,
	\qquad
	\hat\eta=\frac{\Omega_{n+1}r_{0}^{n+1}}{16\pi G_D}\,,
\end{equation}
where $T_{ab}^{(0)}$ is the perfect-fluid stress tensor~\eqref{eq:TabdsSch}.

\subsection{AdS-Schwarzschild-like solutions}
\label{app:firstads}

The purpose of this appendix is to obtain the first-order planar AdS black brane metric in Schwarzschild-like coordinates in radial gauge, starting from the corresponding Ricci-flat black $p$-brane solution and using the AdS/Ricci-flat correspondence as a shortcut~\cite{Caldarelli:2013aaa}. This is convenient both for taking the null limit in the main text (since our bottom-up constructions are also performed in radial gauge) and since we already discussed the Ricci-flat solutions in Appendix~\ref{app:firstflat} above.

Let us begin by briefly recalling the form of AdS/Ricci-flat correspondence, following~\cite{Caldarelli:2013aaa}. A broad class of Ricci-flat metrics in $D=p+n+3$ dimensions can be written as
\begin{equation}
	\label{eq:ricciflatmetric}
	ds_0^2 = \exp\left({\tfrac{2\tilde\phi(r,x;n)}{n+p+1}}\right) 	\left(	d\tilde s_{p+2}^2(r,x;n)+ d\Omega_{n+1}^2	\right)\,,
\end{equation}
where $x$ collectively denotes the $(p+1)$ worldvolume coordinates, and $d\Omega_{n+1}^2$ is the metric on the unit $(n+1)$-sphere. The AdS side is written as
\begin{equation}
	\label{eq:adsmetric}
	ds_\Lambda^2 = d\hat s_{p+2}^2(r,x;d) + \exp\left({\tfrac{2\hat\phi(r,x;d)}{d-p-1}}\right)  d\vec y^{\,2}\,,
\end{equation}
where $d\vec y^{\,2}$ is the metric on a $(d-p-1)$-torus. For the class of solutions relevant here, the correspondence amounts to an analytic continuation in the parameter $n$, with the relation between the metrics given by 
\begin{equation}
	\label{eq:ads-rf-map}
    d = -n\,,\qquad d\hat s_{p+2}^2(r,x;d)=d\tilde s_{p+2}^2(r,x;-d)\,, \qquad \hat\phi(r,x;d)=\tilde\phi(r,x;-d)\,,
\end{equation}
together with a particular reparametrisation of the radial coordinate, which we shall discuss below. For more details, we refer to~\cite{Caldarelli:2013aaa}, which presents the general derivation and discusses the domain of validity.

Our starting point is the first-order Ricci-flat black $p$-brane metric reviewed in Appendix~\ref{app:firstflat}, which may be expressed in Schwarzschild-like coordinates as in~\eqref{eq:dsSchnonnullfirst}. To bring it into the form~\eqref{eq:ricciflatmetric}, we choose
\begin{equation}
	\label{eq:rf-decomposition}
	\tilde\phi=(n+p+1)\ln r\,, 	\qquad 	d\tilde s_{p+2}^2(r,x;r_0;n)=\frac{1}{r^2}\left(ds^2 - r^2 d\Omega_{n+1}^2\right)\,,
\end{equation}
where $ds^2$ on the right-hand side denotes the full Ricci-flat metric (including the $\mathcal{O}(\epsilon)$ correction $f_{\mu\nu}$), and where we have explicitly indicated the dependence on the horizon scale $r_0$. With this identification, \eqref{eq:ricciflatmetric} reproduces \eqref{eq:dsSchnonnullfirst} on the nose.

To obtain a planar AdS black brane metric, it is convenient to invert the Ricci-flat radial coordinate and rename the horizon scale
\begin{equation}
	\label{eq:rf-to-ads-subst}
	n\to -d\,,	\qquad	r \to \frac{1}{r}\,,	\qquad	r_0 \to b\,,
\end{equation}
while the worldvolume coordinates $\sigma^a$ remain unchanged. We also allow for $u^a\to -u^a$ as a convention, but since the ideal-order metric only depends on the product $u_a u_b$, this sign is of no consequence. Under~\eqref{eq:rf-to-ads-subst}, the Ricci-flat factor $r_0^n/r^n$ maps precisely to the familiar AdS blackening factor
\begin{equation}
	\label{eq:blackening-map}
	\frac{r_0^n}{r^n} 	\longrightarrow 	\frac{1}{(br)^d}\,,	\qquad 	f(br):=1-\frac{1}{(br)^d}\,.
\end{equation}
Applying~\eqref{eq:ads-rf-map} and~\eqref{eq:rf-to-ads-subst} then yields an AdS metric of the form
\begin{equation}
	\label{eq:ads-from-rf-intermediate}
	ds_\Lambda^2 =r^2\left(\eta_{ab}+\frac{1}{(br)^d}u_a u_b\right)d\sigma^a d\sigma^b	+	\frac{dr^2}{r^2 f(br)}	+	r^2 d\vec y^{\,2}	+	r^2 f_{\mu\nu}^{\rm (RF)}\Big(\frac{1}{r};\, b;\, -d\Big) dx^\mu dx^\nu\,,
\end{equation}
where $f_{\mu\nu}^{\rm (RF)}$ denotes the Ricci-flat first-order correction discussed in Appendix~\ref{app:firstflat}. Using~\eqref{eq:blackening-map}, the ideal-order part can be rewritten in the standard boosted AdS black brane form
\begin{equation}
	\label{eq:transfAdSmetric}
	ds^{2}	=	r^{2}\Big(\eta_{ab}-\big(f(br)-1\big)u_a u_b\Big)d\sigma^a d\sigma^b	+\frac{dr^{2}}{r^{2}f(br)}	+f_{\mu\nu}(r) dx^{\mu}dx^{\nu}\,,
\end{equation}
where the torus directions with metric $d\vec y^{\,2}$ are spectator directions, and the first-order correction $f_{\mu\nu}(r)$ is obtained from the Ricci-flat functions by the substitutions~\eqref{eq:rf-to-ads-subst}, together with the overall factor $r^2$ already appearing in~\eqref{eq:ads-from-rf-intermediate}. The function $f_{\mu\nu}(r)$ has the same tensorial decomposition as in~\eqref{eq:fmunuflat}, but with coefficients given by
\begin{subequations}
\label{eq:fmununonnullads}
\begin{align}
    f_{1}(r)&=\frac{r^2 b \left(2-(2-d) \left(\frac{1}{rb}\right)^d\right) \log \left(f(br)\right)}{(1-d) d}\, , \\
    f_{2}(r)&=\frac{2 r^2 b \log \left(f(br)\right)}{d}\, , \label{eq:fra-AdS} \\
    f_{3}(r)&=-\frac{d^2 \left(\frac{1}{r}\right)^d \, _2F_1\left(1,\frac{1}{d};1+\frac{1}{d};\left(\frac{1}{rb}\right)^d\right)+r b \left(\left(b^d+(d-1) \left(\frac{1}{r}\right)^d\right) \log \left(f(br)\right)-d \left(\frac{1}{r}\right)^d\right)}{(d-1) d
   r \left(\left(\frac{1}{r}\right)^d-b^d\right)}\, , \\
    f_{4}(r)&=-\frac{b^d \left(\, _2F_1\left(1,\frac{1}{d};1+\frac{1}{d};\left(\frac{1}{rb}\right)^d\right)-1\right)}{r \left(b^d-\left(\frac{1}{r}\right)^d\right)}\, , \\
    f_{5}(r)&=-\frac{\left(\frac{1}{r}\right)^{d+2} b^{1-d} \left(2-\log \left(f(br)\right)\right)}{(1-d) \left(f(br)\right)^2}\, .
\end{align}
\end{subequations}
When $f_{\mu\nu} = 0$ in~\eqref{eq:transfAdSmetric}, we recover the metric of a boosted AdS black brane in Schwarzschild-like coordinates. Moreover, we have explicitly checked that~\eqref{eq:transfAdSmetric} solves the Einstein equations with negative cosmological constant to first order in the derivative expansion, and that the induced boundary energy-momentum tensor is that of a viscous conformal fluid
\begin{equation}
	\label{eq:Tab1AdSnonnull}
	T_{ab}=T_{ab}^{(0)}-2\hat\eta\,\sigma_{ab}\,, 	\qquad 	\hat\eta=\frac{s}{4\pi}\,,
\end{equation}
with $T_{ab}^{(0)}$ and $s$ given in~\eqref{eq:TabAdSideal} and~\eqref{eq:TsAdS}, respectively. This implies that $\hat\eta/s = 1/(4\pi)$, as was famously first obtained holographically in~\cite{Kovtun:2004de}. Finally, we remark that the metric has the expected asymptotically locally AdS falloffs in radial gauge.

\section{Einstein equations for $f_{\mu\nu}$}
\label{app:EFEs-for-fmunu}

In this appendix, we will collect the full expressions for the Einstein equations that we solve for $f_{\mu\nu}$ in the main text. We will do so first for the asymptotically AdS solutions, followed by the asymptotically flat solutions. We  remind the reader that the equations presented here have been explicitly obtained and solved for $d\leq5$ and $(p,n)\in\{1,2,3,4\}$ in the case of asymptotically AdS and asymptotically flat spacetimes, respectively. The general equations found here have been extrapolated from the expressions encountered in our explicit calculation. 

\subsection{AdS-pp-waves}
\label{app:explicitEFE}
In AdS, the simplest equations arise from the off-diagonal spatial components $E_{ij}$ with $i\neq j$, which for general $d>3$ are given by 
\begin{equation}
\label{eq:E_ij}
    2 (d - 2)f_{ij} - (d - 3)r f_{ij}'-r^{2} f_{ij}'' = 0\, .
\end{equation}
The solution for $f_{ij}$ is presented in~\eqref{eq:solfijads}. The next-simplest set of differential equations are $E_{ti} = 0$ and $E_{iz} = 0$, which are coupled and for general $d\geq3$ are given by \begin{equation}
\label{eq:E_ti-E_iz}
\begin{split}
    2d\kappa^{d} f_{iz}+dr\kappa^{d}\big(f_{ti}'-f_{iz}'\big)- (d-3)r^{d+1} f_{ti}'-r^{d+2} f_{ti}''+ 2\left((d-2)r^{d} -d\kappa^{d}\right)f_{ti}&=0 \, , \\
    2d\kappa^{d}f_{ti}+dr\kappa^{d}\big(f_{iz}'-f_{ti}'\big)+(d-3)r^{d+1}f_{i z}'+r^{d+2}f_{i z}''-2\big((d-2)r^{d}+d\kappa^{d}\big)f_{iz}&=0\, ,
\end{split}
\end{equation} 
respectively. The solutions to these equations are presented in~\eqref{eq:solftifziads}. For $d>3$, the difference $E_{ii}-E_{jj}$ gives rise to the equation
\begin{equation}
\label{eq:E_ii-E_jj}
    2(d-2) (f_{ii}-f_{jj})-r(d-3)(f_{ii}'-f_{jj}')-r^2(f_{ii}''-f_{jj}'') =0\,,
\end{equation}
whose solution appears in~\eqref{eq:solf_ii}. The form of the remaining equations $E_{tt} = 0$, $E_{tz} = 0$, $E_{zz}=0$, $E_{ii} = 0$, and $E_{rr} = 0$ depends on the dimension $d$, and we provide them for $d=4$ below as an example
\begin{equation}
\label{eq:EttEtz}
\begin{split}
    0&=-r \Big[r \left(8 \kappa_0^4 \delta^{ij}c_{ij}^{(1)}+\kappa_0^8 f_{tt}''+2 \kappa_0^4 r^4 f_{tz}''+(r^8+\kappa_0^8)f_{zz}''+\left(r^7+2 \kappa_0^4 r^3\right) f_{zz}'+2 r^8 \chi '' +2 r^7 \chi ' \right) 
    \\ & \quad \, + 5 \kappa_0^4 r^8
   f_{rr}'+6 \kappa_0^8 f_{tz}'\Big] +2 r^8 f_{rr} \left(9 r^4-11 \kappa_0^4\right)-12 \kappa_0^8 f_{tt}+24 \kappa_0^8 f_{tz} -16 \kappa_0^4 r^4 \chi 
   \\ & \quad\, +r \left(3 r^{12} f_{rr}'+3 \kappa_0^8 f_{tt}'+2 \kappa_0^4 \left(\kappa_0^4 r f_{tz}''+r^4 f_{tz}'+r^5 f_{zz}''+r^5 \chi '' +3 r^4 \chi ' \right) +3 \kappa_0^8 f_{zz}'+8 r^7
   \chi \right)
   \\ 
   & \quad\, +4 f_{zz} \left(r^8-3 \kappa_0^8\right)\, , \\
   0&=-\kappa_0^4 r \left(8 r^5 \delta^{ij}c_{ij}^{(2)}+\left(3 \kappa_0^4+r^4\right) f_{tt}'+2 \kappa_0^4 r f_{tz}''+r^5 f_{zz}''+3 \kappa_0^4 f_{zz}'+2 r^5
   \chi '' +6 r^4 \chi ' \right) \\ & \quad + 8 \kappa_0^4 r^2 \left(\delta^{ij}c_{ij}^{(1)}+r^4 \delta^{ij}c_{ij}^{(2)}\right) +5 \kappa_0^4 r^9 f_{rr}'+22 \kappa_0^4 r^8 f_{rr}+(\kappa_0^4 r^6 +\kappa_0^8 r^2) f_{tt}''+12 \kappa_0^8 f_{tt}+r^{10} f_{tz}'' \\ 
   & \quad\, +r^9 f_{tz}'+6 \kappa_0^8 r f_{tz}'-4
   f_{tz} \left(6 \kappa_0^8+r^8\right)+\kappa_0^8 r^2 f_{zz}''+\kappa_0^4 r^5 f_{zz}'+12 \kappa_0^8 f_{zz}+16 \kappa_0^4 r^4 \chi \, , \\
   0&=8 \kappa_0^4 r^2 \delta^{ij}c_{ij}^{(1)}+3 r^{13} f_{rr}'+5 \kappa_0^4 r^9 f_{rr}'+2 r^8 f_{rr} \left(11 \kappa_0^4+9 r^4\right)+(r^{10}+2 \kappa_0^4 r^6 +\kappa_0^8 r^2) f_{tt}'' \\ 
   & \quad\, -r \left(\left(3 \kappa_0^8+2 \kappa_0^4 r^4\right) f_{tt}'+2 r \left(\kappa_0^4+r^4\right) \left(\kappa_0^4 f_{tz}''+r^4 \chi '' \right)+3 \kappa_0^8 f_{zz}'+2 \left(r^8+3 \kappa_0^4 r^4\right) \chi ' \right)\\ & 
   \quad\, +r^9 f_{tt}'-4 f_{tt} \left(r^8-3 \kappa_0^8\right)+2( \kappa_0^4 r^5 +3 \kappa_0^8 r) f_{tz}'-24 \kappa_0^8 f_{tz}+\kappa_0^8 r^2 f_{zz}''+12 \kappa_0^8 f_{zz}\\ &
   \quad\, +8 \chi  \left(r^8+2 \kappa_0^4 r^4\right)\, ,
   \\
   0&=r \left(3 r^8 f_{rr}'+(r^5 +\kappa_0^4 r) f_{tt}''+r^4 f_{tt}'+10 \kappa_0^4 f_{tz}'+\kappa_0^4 r f_{zz}''+4 r^3 \chi \right)+18 r^8 f_{rr}
   \\ &
   \quad\, -r \left(5 \kappa_0^4 f_{tt}'+2 \kappa_0^4 r f_{tz}''+r^5
   f_{zz}''+r^4 \left(f_{zz}'+\chi ' \right)+5 \kappa_0^4 f_{zz}'+r^5 \chi '' \right)-4 f_{tt} \left(r^4-2 \kappa_0^4\right)\\ & 
   \quad\, -16 \kappa_0^4 f_{tz}+4 f_{zz} \left(2 \kappa_0^4+r^4\right)=0\, , \\
   0&=12 r^2 \delta^{ij}c_{ij}^{(1)}+12 r^8 f_{rr}+r \left(\kappa_0^4+3 r^4\right) f_{tt}'-2 f_{tt} \left(7 \kappa_0^4+3 r^4\right)-2 \kappa_0^4 r f_{tz}'+28 \kappa_0^4 f_{tz}-3 r^5 f_{zz}'\\ 
   &  \quad\, +\kappa_0^4 r f_{zz}'+6 r^4
   f_{zz}-14 \kappa_0^4 f_{zz}-6 r^5 \chi ' +12 r^4 \chi \, .
\end{split}
\end{equation} 
The solution to these differential equations is given in~\eqref{eq:solttzzrrads}.

\subsection{pp-waves}
\label{app:dynamicalflat}

We now collect the vacuum Einstein equations, $R_{\mu\nu}=0$, used in our discussion about asymptotically flat solutions in Section~\ref{sec:firstflat}. We will derive the equations for general $n = d - 2 - p$, where $d$ is the spatial dimension of the bulk, and $(p+1)$ and $(n+2)$ are the dimensions of the worldvolume and the space transverse to the worldvolume, respectively (cf.~Section~\ref{sec:idealflat}). The simplest of these is $R_{ij} = 0$\footnote{Note that this equation is absent for $p=1$.}
\begin{equation}
\label{eq:E_ijflat}
    (n+1)f_{ij}'+rf_{ij}''=0\,.
\end{equation} 
The solution to this differential equation is given in \eqref{eq:solfij}. The equations $R_{ti}=0$ and $R_{zi}=0$ are coupled and read
\begin{equation}
\label{eq:RtiRzi}
\begin{split}
    ((n+1)r^{n}-n\kappa^n)f'_{ti}+n\kappa^nf'_{iz}+r^{n+1}f''_{ti}&=0\,,
    \\    ((n+1)r^{n}+n\kappa^n)f'_{iz}-n\kappa^nf'_{ti}+r^{n+1}f''_{iz}&=0\,.
\end{split}
\end{equation} 
The solutions appears in~\eqref{eq:solftifzi}. Finally, the equations $R_{tt}=0,\ R_{tz}=0,\ R_{zz}=0$, and $R_{rr}=0$ are given by
\begin{equation}
\label{eq:tt}
\begin{split}
    0&=r^{n+1}\left(\left(-3\kappa^n+\frac{2(n+1)}{n}r^{n}\right)f'_{tt}+\kappa^n(4f'_{tz}-\delta^{ij}f_{ij}'-f_{zz}')+\frac{2}{n}r^{n+1}f_{tt}''\right)\\ & \quad + \kappa^nr^{n+1}f_{rr}'+\kappa^{2n}\left(n(f_{tt}+f_{zz}-2f_{tz})+r(f'_{tt}+f'_{zz}-2f'_{tz})\right)\, , \\
    0&=r^{n+1}\left((n+1)r^{n}f_{tz}'-\kappa^n(f'_{tt}-\delta^{ij}f_{ij}'+f_{zz}')+\frac{2}{n}r^{n+1}f_{tz}''\right)\\ & \quad + \kappa^nr^{n+1}f_{rr}' + \kappa^{2n}\left(n(f_{tt}+f_{zz}-2f_{tz})-r(f'_{tt}+f'_{zz}-2f'_{tz})\right)\, , \\ 
    0&=r^{n+1}\left(\left(-3\kappa^n+\frac{2(n+1)}{n}r^{n}\right)f'_{zz}-\kappa^{n}(4f'_{tz}+\delta^{ij}f_{ij}'-f_{tt}')+\frac{2}{n}r^{n+1}f_{zz}''\right)\\ & \quad + \kappa^nr^{n+1}f_{rr}'+\kappa^{2n}\left(n(f_{tt}+f_{zz}-2f_{tz})+r(f'_{tt}+f'_{zz}-2f'_{tz})\right)\, , \\
    0&=n(n+1)\kappa^{n}\left(f_{tt}+f_{zz}\right)+(n+1)r^{n+1}f'_{rr}+r^{n+2}f''_{tt}+r\kappa^{n}\left(2nf'_{tz}+r\left(f''_{tt}+f''_{zz}\right)\right) \\ & \quad -2n(n+1)\kappa^{n}f_{tz}-r\left(n\kappa^{n}f'_{tt}+n\kappa^{n}f'_{zz}+2r\kappa^{n}f''_{tz}+r^{n+1}\delta^{ij}f_{ij}''\right)\, .
\end{split}
\end{equation}
These are solved in \eqref{eq:solftzr}.

\section{Frame and coordinate transformations}
\label{app:frameandcoord}

Below we gather some useful coordinate and frame transformations used in the main text to study null limits of first-order black branes.

\subsection{Landau vs.~non-thermodynamic frames}
\label{app:frames}

As explained in Section~\ref{sec:nullhydrodynamics}, the null limit of relativistic hydrodynamics depends on the timelike hydrodynamic frame chosen before taking the limit. In the main text we implemented the relevant frame choices directly at the level of the boundary stress tensor. In this appendix we show how the same frame redefinitions arise purely from the bulk viewpoint. It will turn out that they correspond to the freedom to add certain homogeneous first-derivative deformations of the fluid/gravity metric, i.e., to perform a first-order reparametrisation of the collective fields that preserves the gauge choice and maps between different hydrodynamic frames.

The discussion mimics that for AdS black branes~\cite{Bhattacharyya:2007vjd} and for asymptotically flat black branes~\cite{Camps:2010br}: in both of these cases, we start from the standard first-order fluid/gravity solution in ingoing EF coordinates, where the horizon is manifestly regular, and consider its variation under a first-order change the of the collective fields. 

Let $g_{\mu\nu}^{(0)}(u,T)$ denote the ideal-order metric, and $g_{\mu\nu}^{(1)}(\partial u,\partial T)$ the particular first-derivative correction chosen in~\cite{Camps:2010br,Bhattacharyya:2007vjd}. A general first-order perturbation around the same ideal background can be written schematically as
\begin{equation}
	\label{eq:pertmetric}
	g_{\mu\nu} 	=
	g_{\mu\nu}^{(0)}(u,T) 	+ \epsilon\Big( g_{\mu\nu}^{(1)}(\partial u,\partial T) +\delta g_{\mu\nu}(\delta u,\delta T)\Big) +\mathcal{O}(\epsilon^2)\,,
\end{equation}
where we remind the reader that $\epsilon$ is the derivative-counting parameter, and where
\begin{equation}
\label{eq:variation-of-g}
    \delta g_{\mu\nu}(\delta u,\delta T) 	= 	\frac{\partial g_{\mu\nu}^{(0)}}{\partial u^a}\delta u^a 	+  	\frac{\partial g_{\mu\nu}^{(0)}}{\partial T}\delta T\,,
\end{equation}
is the variation induced by a first-order redefinition of the collective fields. Equivalently, one may use $(r_0,u_a)$ for the asymptotically flat case. and $(b,u_a)$ for the AdS case, related to $T$ as in the main text. Linearising around an equilibrium configuration with slow worldvolume dependence, we consider
\begin{equation}
    \label{eq:framemetrictransf}
	u^a=(1,\epsilon(\sigma^b\partial_b u^i+\delta u^i))\,,
	\qquad
	\begin{cases}
		r_0=r_0+\epsilon(\sigma^a\partial_a r_0+\delta r_0)\,, & \text{flat}\,,\\[2pt]
		b=b+\epsilon(\sigma^a\partial_a b+\delta b)\,, & \text{AdS}\,,
	\end{cases}
\end{equation}
with $u_a \delta u^a=0$ so that $\delta u^t=0$, and $\delta u^a,\,\delta r_0 \sim \mathcal{O}(\epsilon)$. A priori, shifting $(u,T)$ changes the metric components in a way that could spoil the gauge choice of the original construction. The key point is that the particular redefinitions we need can be chosen so that the metric remains within the same gauge class. For the asymptotically flat black branes of~\cite{Camps:2010br}, the EF conditions imply that under the variation~\eqref{eq:variation-of-g}, the $rr$- and $\mu\Omega$-components remain unchanged, $\delta g_{rr} = 0 = \delta g_{\mu\Omega}$, implying that the radial and sphere gauge conditions are manifestly preserved. Similarly, for the AdS fluid/gravity solutions of~\cite{Bhattacharyya:2007vjd}, one finds that $\delta g_{rr}=0$. Moreover $\delta g_{ar}$ is proportional to $\delta u_a$, so that the perturbed metric still has the standard EF structure $g_{ar}\propto u'_a$ with $u'_a=u_a+\delta u_a$; this is simply the statement that the velocity entering the EF ansatz has been shifted. Hence, the gauge used in~\cite{Bhattacharyya:2007vjd} is preserved as well. In particular, because \eqref{eq:pertmetric} differs from the original solution only by a first-order parametric deformation of the ideal data, it continues to solve the Einstein equations to first order, provided the usual fluid/gravity constraints are imposed. The remaining freedom is fixed by demanding that the boundary energy-momentum tensor extracted from the perturbed metric takes the desired hydrodynamic form. To achieve this, we compute the Brown--York quasilocal energy-momentum tensor of the metric~\eqref{eq:pertmetric} using the prescription described in the main text (cf.~\eqref{eq:AdSTabdefn} and~\eqref{eq:BYflat}) and demand that it matches either:
\begin{enumerate}
    \item[(i)] for the asymptotically flat case, the non-thermodynamic form~\eqref{eq:nullLandauT1}, and
    \item[(ii)] for the AdS case, the non-thermodynamic form~\eqref{eq:newnonthermoT}.
\end{enumerate}
Imposing this matching uniquely fixes the first-order field redefinitions. One finds:
\begin{align}
	\label{eq:frame-shift-flat}
	\text{flat:}\qquad 	\delta u_a&=\frac{r_0}{n} \dot u_a\,,
	&
	\delta r_0&=\frac{2}{n^2(n+1)} r_0^2 \theta\,,
	\\[4pt]
	\label{eq:frame-shift-ads}
	\text{AdS:}\qquad
	\delta u_a&=\frac{2b \dot u_a-\partial_a b-u_a u^c\partial_c b}{2d}\,,
	&
	\delta b&=\frac{2b^2\theta-(d-1)b u^c\partial_c b}{d^2(d-1)}\,.
\end{align}
These are precisely the frame redefinitions obtained from the boundary hydrodynamic analysis in \eqref{eq:frametransfflat} and \eqref{eq:hydroframeads}, respectively.

In the asymptotically flat case it is often convenient to extract the effective energy-momentum tensor in a Schwarzschild-like gauge,  as we did in the main text, rather than directly in EF coordinates. When starting from the EF form, one may therefore first perform the standard first-order coordinate transformation, with the fields defined as in \eqref{eq:framemetrictransf}, to the Schwarzschild-like radial gauge before evaluating the Brown--York tensor. This is a technical step and does not affect the logic: the matching of energy-momentum tensors fixes the same redefinitions~\eqref{eq:frame-shift-flat}.

Finally, plugging \eqref{eq:frame-shift-flat} and \eqref{eq:frame-shift-ads} into the corresponding first-order bulk metrics produces new first-order solutions whose corresponding energy-momentum tensors are exactly \eqref{eq:nullLandauT1} (flat) and \eqref{eq:newnonthermoT} (AdS). This establishes the bulk origin of the hydrodynamic frame freedom.

\subsection{Coordinate transformation for finite null limit}
\label{app:fraremoval}

When taking the null limit of the fluid metrics for both asymptotically flat spacetimes and AdS, the component $f_{ra}$ (cf.~Eqs.~\eqref{eq:fra-flat} and~\eqref{eq:fra-AdS}) diverges in $r_0^{-1}$ and $b$, respectively. For example, the first term in the expression for $f_{ra}$ for AdS, given in~\eqref{eq:fra-AdS}, behaves as
\begin{equation}
    \dot{u}_af_2\sim f_2b^{d(2+\alpha)/2}U^b\partial_bU_a\rightsquigarrow\infty\,,
\end{equation}
for $\alpha=-(d+2)/d$ and $\alpha=(d-2)/d$ used in the Landau and non-thermodynamic frames, respectively. Of course, it is always possible to choose a value $|\tilde \alpha|>|\alpha|$ such that the limit is well-defined, but this would imply that all the other components of the metric go to zero.
Furthermore, the bottom-up construction shows that $f_{ra}=\mathcal{O}(\epsilon^2)$, where $\epsilon$ is the formal parameter that counts the number of worldvolume derivatives introduced in~\eqref{eq:vkcollvar}. This appendix demonstrates the existence of a coordinate transformation that sets $f_{ra} = 0$ up to second order in derivatives, while preserving the form of the remaining components of the fluid metric up to that same order. 

Consider an infinitesimal change of coordinates parametrised by $\xi^\mu$
\begin{equation}
\label{eq:inf-diffeo}
    x^\mu \to x'^\mu = x^\mu + \epsilon \xi^\mu(x)\,.
\end{equation}
The fluid metric admits a derivative expansion of the form
\begin{equation}
    g_{\mu\nu} = g^{(0)}_{\mu\nu}(r) + \epsilon g^{(1)}_{\mu\nu}(r,\sigma) + \mathcal{O}(\epsilon^2)\,,
\end{equation}
where $g^{(0)}_{\mu\nu}(r)$ is the background metric, which satisfies $g^{(0)}_{ra} = 0$, which in turns implies that 
\begin{equation}
\epsilon g^{(1)}_{ra}(r) = f_{ra}(r)\,,   
\end{equation}
is a function of $r$ only. Under the coordinate transformation~\eqref{eq:inf-diffeo}, the fluid metric transforms as
\begin{equation}
\label{eq:metric-trafo}
    g'_{\mu\nu} = g_{\mu\nu} - \epsilon \pounds_\xi g^{(0)}_{\mu\nu} + \mathcal{O}(\epsilon^2)\,,
\end{equation}
so that, in particular, 
\begin{equation}
\label{eq:component-trafos}
    g'^{(0)}_{\mu\nu} = g^{(0)}_{\mu\nu}\,,\qquad g'^{(1)}_{\mu\nu} = g^{(1)}_{\mu\nu} - \pounds_\xi g^{(0)}_{\mu\nu}\,.
\end{equation}
Next, we choose the parameter $\xi^\mu$ to only have legs in the worldvolume directions and demand that it does not depend on any transverse directions, should they be present; that is,
\begin{equation}
    \xi^\mu = \xi^a (r,\sigma) \delta^\mu_a\,. 
\end{equation}
This implies that
\begin{equation}
    \pounds_\xi g^{(0)}_{r a} = g^{(0)}_{ab}\D_r \xi^b\,,
\end{equation}
and therefore, using~\eqref{eq:component-trafos}, we can set $g'_{\mu\nu} = $ if we choose $\xi^a$ such that $g^{(1)}_{\mu\nu} - \pounds_\xi g^{(0)}_{\mu\nu} = \mathcal{O}(\epsilon)$, translating into the requirement
\begin{equation}
    \D_r \xi^a = g^{(0)ab}(r)g^{(1)}_{rb}(r) + \mathcal{O}(\epsilon)\,,
\end{equation}
which we can integrate to get
\begin{equation}
\label{eq:Xi-a}
    \xi^a(r,\sigma) = c^a_r(\sigma) + \Xi^a(r)\,,\qquad \Xi^a(r):=\int_{r_*}^r dr'\, g^{(0)ab}(r')g^{(1)}_{rb}(r')\,,
\end{equation}
where $c^a_r(\sigma)$ is an integration constant, and $r_*$ is a reference radius parametrising the integration constant $c^a_r(\sigma) = \xi^a(r_*,\sigma)$. We will not need explicit expressions for $\xi^a$, which depend on the dimension, the choice of $r_*$, and whether we consider asymptotically flat or AdS backgrounds. In particular, as advertised, this coordinate transformation leaves the fluid metric unchanged up to $\mathcal{O}(\epsilon^2)$, which follows from~\eqref{eq:metric-trafo} and the fact that $g^{(0)}_{\mu\nu}$ does not depend on the worldvolume coordinates $\sigma^a$:
\begin{equation}
    \delta g_{rr} = \mathcal{O}(\epsilon^2)\qquad\text{and}\qquad \delta g_{ab} = \mathcal{O}(\epsilon^2)\,.
\end{equation}
This shows that we can always change coordinates such that $f_{ra} = \mathcal{O}(\epsilon^2)$, while preserving the form of the other components of the fluid metric up to second order in derivatives. We have explicitly checked that the metrics obtained via this procedure solve Einstein's equations in a hydrodynamic expansion. This was done for both asymptotically flat and AdS spacetimes, and in both the Landau and non-thermodynamic frames. 

Finally, we now discuss how a residual radial coordinate transformation allows to get rid of the function $\chi(r)$ that appears in various components of $f_{\mu\nu}$ (cf.~\eqref{eq:solf_ii} and~\eqref{eq:solttzzrrads}). As explained in Section~\ref{sec:AdSbottomup}, $\chi(r)$ drops out of all calculations, and so we set it to zero. As we now demonstrate, this can be achieved by performing a residual radial coordinate transformation, i.e., a diffeomorphism that preserves the condition $g_{ra} = \mathcal{O}(\epsilon^2)$. Working infinitesimally, as above, this amounts to requiring that
\begin{equation}
    0 \overset{!}{=} \delta_\zeta g_{ra} = -\epsilon \pounds_\zeta g^{(0)}_{ra} = - \epsilon( r^{-2}\D_a \zeta^r + \D_r\zeta_b)\,,
\end{equation}
where $\zeta^\mu$ is the generator of infinitesimal residual diffeomorphisms preserving radial gauge, and where we used that $g_{rr}^{(0)} = r^{-2}$. In particular, $\tilde\zeta^\mu = (\tilde\zeta^r(r),0)$ generates a residual diffeomorphism; focusing on the radial component $g_{rr}$, $\tilde\zeta$-transformations act as
\begin{equation}
    \delta_{\tilde\zeta}g_{rr} = -\epsilon \pounds_{\tilde\zeta}g_{rr} = \frac{2\epsilon}{r^2}\left( \frac{\tilde\zeta^r}{r} - \D_r\tilde{\zeta}^r \right)\,,
\end{equation}
up to seond order in $\epsilon$. Comparing with the expression for $f_{rr}$ in~\eqref{eq:solttzzrrads}, we see that identifying
\begin{equation}
    \epsilon \zeta^r(r) = \frac{\chi(r)}{2r}\,,
\end{equation}
precisely cancels out all terms involving $\chi$ in $g_{rr}$ up to $\mathcal{O}(\epsilon^2)$. To see that this choice of residual diffeomorphism $\tilde\zeta$ accomplishes the same for the remaining components of $g_{ab}$ that involve $\chi$, note that $\tilde\zeta$ does not depend on $\sigma^a$ implying $\delta g_{ab} = -\epsilon\, \tilde\zeta^r\D_r g^{(0)}_{ab}$. and since
\begin{equation}
    g^{(0)}_{ab} = r^2 \eta_{ab} + r^{2-d}\kappa^d v^{(0)}_a v^{(0)_b}\,,
\end{equation}
we find that
\begin{equation}
    \delta_{\tilde\zeta}g_{ab} = -\chi \eta_{ab} + \frac{d-2}{2}r^{-d}\kappa^d \chi v^{(0)}_a v^{(0)}_b\,,
\end{equation}
which again removes all dependence on $\chi$ in $f_{ab}$ as it appears in~\eqref{eq:solf_ii} and~\eqref{eq:solttzzrrads} after using that $v_a^{(0)} = (1,0_i,1)$. This explicitly shows that the function $\chi(r)$ corresponds to a residual radial coordinate transformation, and explains why it does not feature in the energy-momentum tensor.

\bibliographystyle{utphys2}
\bibliography{bibliography}

\end{document}